\documentclass[a4paper,11pt]{article}
\pdfoutput=1 % if your are submitting a pdflatex (i.e. if you have
\usepackage{graphicx}  % needed for figures
\usepackage{dcolumn}   % needed for some tables
\usepackage{bm}        % for math
\usepackage{amssymb}   % for math
\usepackage[latin1]{inputenc}
\usepackage[english]{babel}
\usepackage{shapepar}
\usepackage{jcappub} 
\usepackage[T1]{fontenc} % if needed
\usepackage{amsmath}
\usepackage{amsthm}
\usepackage{array}
\usepackage{fancybox}
\usepackage{multirow}
\usepackage{appendix}
\usepackage{epsfig}
\usepackage{amsmath}
\usepackage{amsthm}
\usepackage{array}
\usepackage{fancybox}
\usepackage{multirow}
\usepackage{appendix}
\usepackage{epsfig}
\usepackage{cases}
\usepackage{forest}
\DeclareMathOperator{\sech}{sech}

\theoremstyle{definition}

\begin{document}

%\pacs{}
%\maketitle

\title{\boldmath Focusing conditions for extended teleparallel gravity theories}

\author[a]{U. K. Beckering Vinckers,}\emailAdd{bckulr002@myuct.ac.za}
\author[a]{\'A. de la Cruz-Dombriz,}
\author[a,b]{F. J. Maldonado Torralba}

\affiliation[a]{Cosmology and Gravity Group, Department of Mathematics and Applied Mathematics, University of Cape Town, Rondebosch 7701, Cape Town, South Africa.}
\affiliation[b]{Van Swinderen Institute, University of Groningen, 9747 AG Groningen, The Netherlands.}

%\emailAdd{f.j.maldonado.torralba@rug.nl}

\abstract{In the context of extended theories of teleparallel gravity $f(T)$ we derive the focusing conditions for a one-parameter dependent congruence of timelike auto-parallels of the Levi-Civita connection. 
We also consider the $f(T)$ field equations for a general metric tensor before moving on to consider a spatially flat Robertson-Walker space-time. Following this, we study the expansion rate for a one-parameter dependent congruence of timelike auto-parallel curves of the Levi-Civita connection. Given the fact that test particles follow auto-parallels of the Levi-Civita connection, the torsion-free Raychaudhuri equation is used in order to determine the desired focusing conditions. Finally we study the obtained focusing conditions for three $f(T)$ paradigmatic cosmological models and discuss the satisfaction or violation of these conditions. Through this, we find $f(T)$ models that allow for the weak and strong focusing conditions to be satisfied or violated. It is mentioned that this behaviour can also be found in the so-called $f(R)$ and $f(Q)$ theories.}

\maketitle

%%%%%%%%%%%%%%%%%%%%%%%%%%%
%%%%%%%%%%%%%%%%%%%%%%%%%%%

%%%%%%%%%%%%%%%%%%%%%%%%
\section{Introduction}
\label{intro}
%\subsection{General concepts}
It is well known that General Relativity (GR), which provides us with a description of space, time and gravity, has been usually formulated with symmetric connections only \cite{wald4}. More specifically, GR is formulated on the so-called Riemannian space-time whose natural choice of an affine connection is the unique torsion-free and metric compatible connection, known as the Levi-Civita connection. When one considers the convergence or divergence of neighbouring curves in torsion-free theories of gravity, such as GR, one often turns to the Raychaudhuri equation \cite{wald4,raychaudhuri}. The Raychaudhuri equation, which is known for playing a significant role in the proofs of the singularity theorems \cite{wald4,ellis,raychaudhuri,hawking}, can be derived through the use of the Levi-Civita connection for a congruence of curves.

In the present work, we consider theories of gravity for which we have non-symmetric connections, and consequently torsion. In particular, we consider theories constructed on the Riemann-Cartan space-time whose natural choice of an affine connection is not necessarily symmetric while being metric compatible \cite{tseng}. For such space-times, the Levi-Civita connection is not, in general, the most natural choice of an affine connection and is instead the so-called Cartan connection \cite{tseng,audretsch}. When studying Riemann-Cartan space-times, however, it is still possible to consider the Levi-Civita connection as an additional structure \cite{kobayashi}. It is, therefore, possible to still consider the convergence or divergence of curves generated by the Levi-Civita connection in general Riemann-Cartan space-times. 
The convergence or divergence of such curves can be described by the aforementioned Raychaudhuri equation which we shall refer to as the \textit{torsion-free Raychaudhuri equation} in the present work. Given the fact that the Levi-Civita connection is not the natural choice of an affine connection in the Riemann-Cartan space-time, the torsion-free Raychaudhuri equation cannot, in general, be used to describe the convergence or divergence of curves generated by the Cartan connection. Consequently, in order to study a congruence of curves generated by the Cartan connection, we aim at determining a Raychaudhuri equation describing the convergence and divergence of these curves. One such equation can be found in \cite{Capozziello:2001mq,kar} and we refer to this result as the \textit{non-null torsion Raychaudhuri equation}. This equation is derived by decomposing the tensor field obtained by acting the Cartan covariant derivative on the tangent vector field of a curve contained in a congruence. While the non-null torsion Raychaudhuri equation is valid in Riemann-Cartan space-time, the associated kinematic quantities do not necessarily describe the physical expansion, shear and twist. In \cite{Luz:2017ldh,Dey:2017fld}, a Raychaudhuri equation involving torsion and whose kinematic quantities give the physical expansion, shear and twist is derived. In this note, we refer to this result as the \textit{geometric Raychaudhuri equation}. In \cite{mena}, the geometric Raychaudhuri equation is made use of in studying singularity theorems in affine theories of gravity with torsion. We note that the more general Raychaudhuri equation for the case of a space-time whose natural choice of an affine connection is not metric compatible in addition to having a non-vanishing torsion tensor is derived in \cite{iosifidis}.

Once we have derived the non-null torsion and geometric Raychaudhuri equations, we will move on to discuss the implication of such key equations in the context of $f(T)$ theories of gravity. In such theories, the underlying Riemann-Cartan space-time requires the non-null torsion Riemannian curvature to vanish. Such a space-time, which is referred to as the Weitzenb\"ock space-time, is constructed by requiring that the natural choice of an affine connection is the so-called Weitzenb\"ock connection \cite{weitz}. The $f(T)$ theories of gravity, where $T$ is the torsion scalar, is a modification of the so-called Teleparallel Equivalent of General Relativity (TEGR) \cite{cai2015,bengochea}. The reason TEGR is equivalent to GR is that, apart from a surface term that has no contribution, the TEGR action is exactly the Einstein-Hilbert action \cite{cai2015}. The TEGR theory of gravity was first introduced by Einstein and the interested reader is directed to the work \cite{unzicker} which contains the first English translations of Einstein's original papers on the subject. The formulation of TEGR through the construction of a gravitational Lagrangian was first done in \cite{maluf} and later in \cite{deandrade}. While TEGR is a Lorentz invariant theory, the field equations of a general $f(T)$ theory are not necessarily Lorentz invariant \cite{tamanini}. The fact that these theories display this property has lead to recent interest \cite{alvaro,tamanini,mengwang,ferraro}. In particular, the validity of Birkhoff's Theorem in $f(T)$ theories is discussed in \cite{mengwang,tamanini,ferraro}. It was argued in \cite{mengwang} that the first inference of Birkhoff's theorem, which states that a spherically symmetric metric for which a vacuum is considered is static, holds in $f(T)$ theories of gravity. The authors of \cite{mengwang} made this argument by starting with a spherically symmetric metric and then constructing a diagonal tetrad. As pointed out in \cite{tamanini,ferraro}, the diagonal tetrad used in \cite{mengwang} constrains the general $f(T)$ theory of gravity being considered to be TEGR and thus the argument given in \cite{mengwang} is limited to that case. As per the terminology used in \cite{tamanini}, the diagonal tetrad used in \cite{mengwang} is an example of a "bad tetrad". Further studies conducted in the context of $f(T)$ theories of gravity include the study of junction conditions in these theories \cite{alvaro} as well as constraining specific $f(T)$ theories through the use of Supernovae Type Ia and Baryon Acoustic Oscillation data \cite{nunes2016}. The review article \cite{d'agostino} discusses a model-independent reconstruction of $f(T)$ theories through the use of the so-called cosmographic method. For theories that can be written as a function of the torsion scalar, $T$, a scalar field and its kinetic term, a study of these theories, as well as a comparison with Hubble data, can be found in \cite{Abedi:2018lkr}. The so-called $f(T,B)$ theories, where $B$ is the Levi-Civita divergence of the trace of the torsion tensor, have been studied in \cite{Otalora:2016dxe,Bahamonde:2016grb,Bahamonde:2015zma,Bahamonde:2019jkf,Capozziello:2019msc}. Theories resulting from the inclusion of a Gauss-Bonnet term in the $f(T)$ Lagrangian have also been studied \cite{Kofinas:2014owa,Kofinas:2014daa,Bahamonde:2016kba,Capozziello:2016eaz,delaCruz-Dombriz:2017lvj,delaCruz-Dombriz:2018nvt}. In addition, a topic of interest for the case of TEGR is the status of the theory as a gauge theory of the translation group \cite{fontanini,pereira2019,ledelliou2019}. Finally, it is worth mentioning that recently, the set of solutions of $f(T)$ gravity in the Minkowski metric has been studied \cite{Jimenez:2020ofm}, finding out that a new mode is present at fourth-order perturbation, signalling a strong coupling problem.

In this note, we are concerned with studying the convergence or divergence of paths followed by test particles. By studying such paths as well as by making use of the $f(T)$ field equations, we will be able to determine the so-called focusing conditions imposed in $f(T)$ theories of gravity. The $f(T)$ focusing conditions for the case where fundamental congruences are considered have been derived before in \cite{liu}. In the present work, we derive the focusing conditions given a one-parameter dependent congruence of timelike auto-parallels of the Levi-Civita connection.

This work is organised as follows: 
first in Section \ref{Sec:2} we shall discuss in detail the formulation of the so-called Riemann-Cartan space-time. Herein, we shall also present a thorough discussion on auto-parallels and extremal curves in Riemann-Cartan space-times to be distinguished from the well-known geodesic curves in the usual Riemannian space-time. 
Then in Section \ref{Sec:3} we shall derive the non-null torsion and geometric Raychaudhuri equations.
In Section \ref{Sec:4}, we shall focus our attention on $f(T)$ theories of gravity. Such theories of gravity possess a vanishing non-null torsion Riemann tensor \cite{cai2015}. In order to ensure that property, we shall consider the Weitzenb\"ock space-time by taking the natural choice of an affine connection to be the Weitzenb\"ock connection. 
There, we shall discuss the $f(T)$ field equations for a general metric tensor before specifying a spatially flat Robertson-Walker space-time in Section \ref{Sec:5}. Following this, we study the expansion rate for a one-parameter dependent congruence of timelike auto-parallel curves of the Levi-Civita connection. By noting that test particles follow auto-parallels of the Levi-Civita connection \cite{deandrade,deandrade2}, in Section \ref{Sec:6} we shall turn our attention to the torsion-free Raychaudhuri equation in order to determine the focusing conditions. Such conditions are thoroughly studied in Section \ref{Sec:7} for three paradigmatic $f(T)$ cosmological models and the satisfaction or violation of these conditions shall be discussed there.
Finally, we present our Conclusions in Section \ref{Sec:8}.

{\bf{Notation and conventions}}: In the following we shall make use of natural units, i.e., $c=G_N=1$ and consider four-dimensional space-times with 
a signature $(-+++)$. We shall use Greek indices to denote space-time indices whereas Latin characters would be reserved for tangent space indices. Components of a type $(m,n)$ tensor, say $\bm{B}$, shall be written as $B^{\mu_1\dots \mu_m}_{\ \ \ \ \ \ \ {\nu_1\dots \nu_n}}$. Furthermore, for the tensor $\bm{B}$, we make use of the following notational convention as used in \cite{wald4}
\begin{align}{\label{51}}
B^{\mu_1\dots \mu_m}_{\ \ \ \ \ \ \ {[\nu_1\dots \nu_n]}} = \frac{1}{n!}\sum_\pi \delta_\pi B^{\mu_1\dots \mu_m}_{\ \ \ \ \ \ \ \nu_{\pi(1)}\dots \nu_{\pi(n)}} \ ,
\end{align}
where $\delta_\pi$ is $-1$ for odd permutations and $+1$ for even permutations. To illustrate this notational convention, consider for example the case where the tensor $\bm{B}$ is of type $(0,3)$. Then we write $B_{[\mu|\nu|\beta]} = \frac{1}{2}(B_{\mu\nu\beta} - B_{\beta\nu\mu})$.
In addition, we make use of the notational convention for a type $(m,n)$ tensor $\bm{B}$ \cite{wald4}
\begin{align}
B^{\mu_1\dots \mu_m}_{\ \ \ \ \ \ \ {(\nu_1\dots \nu_n)}} = \frac{1}{n!}\sum_\pi B^{\mu_1\dots \mu_m}_{\ \ \ \ \ \ \ \nu_{\pi(1)}\dots \nu_{\pi(n)}} \ .
\end{align}
Following our example for the case where $\bm{B}$ is a type $(0,3)$ tensor, we write $B_{(\mu|\nu|\beta)} = \frac{1}{2}(B_{\mu\nu\beta} + B_{\beta\nu\mu})$.

%%%%%%%%%%%%%%%%%%%%%%%%%%%%%%%%%%%%%%%%%%%%%%
\section{Space-time and torsion}
\label{Sec:2}

In this Section we shall first address the Riemannian space-time, which we denote as $\mathcal{V}_4$, in \ref{Sec:2:A} before moving on to discuss the Riemann-Cartan space-time, which we denote as $\mathcal{U}_4$, in \ref{Sec:2:B}. Below we shall discuss how certain quantities arise in a torsion-free framework as well as how these quantities may be generalised for gravitational theories with non-null torsion. 
Consequently, the non-Riemannian contributions of the non-null torsion quantities can be obtained simply by subtracting the associated quantities present in the Riemannian space-time $\mathcal{V}_4$ \cite{hehl76}. More specifically, below we shall show how to relate the metric compatible Levi-Civita connection to the metric compatible Cartan connection through the use of the so-called contorsion tensor \cite{hehl76,sl}. In addition, we shall briefly examine how it is possible to relate the torsion-free Riemann tensor that arises in Riemannian space-time to the non-null torsion Riemann tensor that arises in Riemann-Cartan space-time through the use of torsion tensor terms as well as their first order derivatives. 
Also, at the end of this Section we shall revise the notion of geodesics in Riemannian space-time in \ref{Sec:2:C}
and the notion of auto-parallels and extremal curves in the Riemann-Cartan space-time in \ref{Sec:2:D}.

\subsection{The Riemannian space-time $\mathcal{V}_4$}
\label{Sec:2:A}
We begin by considering a four-dimensional, Hausdorff and connected $C^{\infty}-$manifold $\bm{\mathcal{M}}$ endowed with an additional structure, $D$, referred to as an affine connection \cite{ellis}. This extra structure, $D$, is such that, for all dual vector fields $\nu_\beta$, we have \cite{wald4,ellis}
\begin{align}\label{tfcs}
D_\alpha\nu_\beta = \partial_\alpha\nu_\beta - \Gamma^\rho_{\ \alpha\beta}\nu_\rho \ ,
\end{align}
where $\Gamma^\rho_{\ \alpha\beta}$ are connection coefficients. When dealing with the manifold $\bm{\mathcal{M}}$, we wish to consider connections that are necessarily torsion-free. That is, we wish to consider connections that are symmetric in the two lower indices, i.e.,
%\begin{align}
$\Gamma^\rho_{\ \alpha\beta}=\Gamma^\rho_{\ \beta\alpha}$. 
%\end{align}
In addition, given a Lorentz metric tensor, $\bm{g}$, we wish to take the affine connection $D$ to be metric compatible, i.e., $D_\nu g_{\alpha\beta}=0$.  As widely known,
the Levi-Civita connection can be easily obtained from a given metric with the coefficients \cite{wald4}
\begin{align}\label{lccs}
\Gamma^\rho_{\ \alpha\beta} = \frac{1}{2}g^{\rho\sigma}\left(\partial_\alpha g_{\beta\sigma} + \partial_\beta g_{\alpha\sigma} - \partial_\sigma g_{\alpha\beta}\right) \ . 
\end{align}
The Levi-Civita connection is the unique affine connection that is both metric compatible as well as torsion-free \cite{chern}. Following \cite{tseng}, we define the \textit{Riemannian space-time}, denoted as $\mathcal{V}_4$, to be the tuple $(\bm{\mathcal{M}},\bm{g},D)$ where the natural choice of an affine connection, $D$, is the Levi-Civita connection.\\ \\
Before moving on to discuss affine connections in Riemann-Cartan space-times, we wish to first consider the notion of curvature in our Riemannian space-time $\mathcal{V}_4$. As shown in \cite{wald4}, one can show that $\left(D_\alpha D_\beta-D_\beta D_\alpha\right)$ is a linear map and thus, one can introduce a tensor field, $R_{\alpha\beta\sigma}^{ \ \ \ \ \rho}$, that is such that
\begin{align}\label{torfreerie}
\left(D_\alpha D_\beta-D_\beta D_\alpha\right)\nu_\sigma = R_{\alpha\beta\sigma}^{ \ \ \ \ \rho}\nu_\rho \ , 
\end{align}
for all dual vector fields $\nu_\sigma$. The tensor field $R_{\alpha\beta\sigma}^{ \ \ \ \ \rho}$ is referred to as the \textit{torsion-free Riemann tensor}\footnote{Equation (\ref{torfreerie}) can be used to express the torsion-free Riemann tensor in terms of the Christoffel symbols, $\Gamma^\rho_{\ \alpha\beta}$, given in equation (\ref{lccs}) as well as their first order derivatives, $\partial_\sigma\Gamma^\rho_{\ \alpha\beta}$. The resulting expression yields
%\begin{eqnarray}\label{rtcsctf}
$R_{\alpha\beta\sigma}^{ \ \ \ \ \rho}= \partial_\beta{\Gamma}^\rho_{\ \alpha\sigma} - \partial_\alpha{\Gamma}^\rho_{\ \beta\sigma}+ {\Gamma}^\delta_{\ \alpha\sigma}{\Gamma}^\rho_{\ \beta\delta}- {\Gamma}^\delta_{\ \beta\sigma}{\Gamma}^\rho_{\ \alpha\delta}$ \cite{wald4}. 
%\nonumber . 
%\end{eqnarray}
}
and gives us the notion of curvature in our Riemannian space-time $\mathcal{V}_4$.  

The torsion-free Riemann tensor provides us with the notion of curvature in gravitational theories whose natural choice of an affine connection is the Levi-Civita connection such as GR \cite{wald4}. In other words, the torsion-free Riemann tensor provides us with the notion of curvature for gravitational theories that are constructed through the use of the Riemannian space-time $\mathcal{V}_4$. When considering such gravitational theories, one often makes use of the \textit{torsion-free Ricci tensor}, $R_{\alpha\beta}$, defined as $R_{\alpha\beta} := R_{\alpha\rho\beta}^{\ \ \ \ \rho}$ as well as its trace $R:=R_\alpha^{\ \alpha}$ which we shall refer to as the \textit{torsion-free Ricci scalar}. 
%In the following subsection, we will consider space-times that have affine connections that are not necessarily torsion free, however, as we will see, we can still make use of the Levi-Civita connection, $D$, in considering non-null torsion theories of gravity in addition to non-null torsion affine connections. 
%
In what follows, we shall discuss the notion of curvature in gravitational theories for which the underlying space-time is the Riemann-Cartan space-time. In order to achieve this, one must consider an additional curvature tensor, which we shall refer to as the \textit{non-null torsion Riemann tensor}, which completely describes the notion of curvature in Riemann-Cartan space-time.

\subsection{The Riemann-Cartan space-time $\mathcal{U}_4$}
\label{Sec:2:B}
Consider a four-dimensional, Hausdorff and connected $C^\infty-$ manifold $\bm{\mathcal{M'}}$ with affine connection $\nabla$ as an additional structure on $\bm{\mathcal{M'}}$ \cite{ellis}. Here, we wish to consider affine connections that are not necessarily torsion-free, although metric compatible. For every dual vector field $\nu_b$, we have
\begin{align}\label{ft1}
\nabla_\alpha\nu_\beta = \partial_\alpha\nu_\beta - \tilde{\Gamma}^\rho_{\ \alpha\beta}\nu_\rho \ ,
\end{align}
where $\tilde{\Gamma}^\rho_{\ \alpha\beta}$\footnote{Here, we have used the tilde to distinguish between the connection coefficients given in equation (\ref{ft1}) and the connection coefficients given in equation (\ref{tfcs}).} 
are the connection coefficients of the affine connection $\nabla$ \cite{wald4,hehl76,sl}.  In addition, we define the \textit{torsion tensor}, $T^\rho_{\ \alpha\beta}$, to be \cite{wald4,hehl76,sl}
\begin{align}\label{tordef}
T^\rho_{\ \alpha\beta} := \tilde{\Gamma}^\rho_{\ \alpha\beta} - \tilde{\Gamma}^\rho_{\ \beta\alpha} \ .
\end{align}
The fact that the connection coefficients are not necessarily symmetric in their lower indices imply that, given a metric tensor, $\bm{g}$, the Christoffel symbols leading to a metric compatible affine connection are not necessarily those of the Levi-Civita connection given in equation (\ref{lccs}). Instead, the Christoffel symbols of the affine connection $\nabla$ are given by the expression
\begin{align}\label{nntcs}
\tilde{\Gamma}^\rho_{\ \alpha\beta} = \Gamma^\rho_{\ \alpha\beta} + K^\rho_{\ \alpha\beta} \ , 
\end{align}
where $\Gamma^\rho_{\ \alpha\beta}$ are the connection coefficients of the Levi-Civita connection and $K^\rho_{\ \alpha\beta}$ is the so-called \textit{contorsion tensor} which is defined through the following expression \cite{hehl76,sl}
\begin{align}\label{contorsion}
K_{\sigma\alpha\beta} := \frac{1}{2}(T_{\sigma\alpha\beta} + T_{\alpha\sigma\beta} + T_{\beta\sigma\alpha}) \ . 
\end{align}
We refer to the metric compatible affine connection $\nabla$ whose connection coefficients are given by equation (\ref{nntcs}) as the \textit{Cartan connection} \cite{audretsch}. Given a metric tensor $\bm{g}$, we define the \textit{Riemann-Cartan space-time}, denoted as $\mathcal{U}_4$, to be the tuple $\left(\bm{\mathcal{M}'},\bm{g},\nabla\right)$ where $\nabla$ is the Cartan connection \cite{tseng}. Since the Cartan connection is the natural choice of an affine conention in $\mathcal{U}_4$, in the following we shall rely on this connection to yield quantities that describe completely certain aspects of the space-time such as curvature. We note, however, that the Levi-Civita connection is also metric compatible in $\mathcal{U}_4$ and that it is also possible to make use of the Levi-Civita connection as an additional structure on the manifold $\bm{\mathcal{M}'}$. 

We now wish to introduce the notion of curvature for the Riemann-Cartan space-time $\mathcal{U}_4$. Given the fact that the combination $(\nabla_\alpha\nabla_\beta - \nabla_\beta\nabla_\alpha + T^\rho_{\ \alpha\beta}\nabla_\rho)$ is a linear map, we may introduce a tensor field, $\tilde{R}_{\alpha\beta\sigma}^{\ \ \ \ \rho}$, that is such that 
\begin{align}\label{nntrd}
\left(\nabla_\alpha\nabla_\beta - \nabla_\beta\nabla_\alpha + T^\rho_{\ \alpha\beta}\nabla_\rho\right)\nu_\sigma = \tilde{R}_{\alpha\beta\sigma}^{\ \ \ \ \rho}\nu_\rho \ ,
\end{align}
for all dual vector fields $\nu_\beta$ \cite{sl}. We refer to $\tilde{R}_{\alpha\beta\sigma}^{\ \ \ \ \rho}$  as the \textit{non-null torsion Riemann tensor}\footnote
{
By making use of the definition for the non-null torsion Riemann tensor given in equation (\ref{nntrd}) together with equation (\ref{ft1}), it is possible to obtain an expression for $\tilde{R}_{\alpha\beta\sigma}^{\ \ \ \ \rho}$ in terms of the Cartan connection coefficients, $\tilde{\Gamma}^\rho_{\ \alpha\beta}$, as well as their first order derivatives \cite{sl}. This expression is as follows
%\begin{eqnarray}\label{rtcsc}
$\tilde{R}_{\alpha\beta\sigma}^{ \ \ \ \ \rho}= \partial_\beta\tilde{\Gamma}^\rho_{\ \alpha\sigma} - \partial_\alpha\tilde{\Gamma}^\rho_{\ \beta\sigma}+ \tilde{\Gamma}^\delta_{\ \alpha\sigma}\tilde{\Gamma}^\rho_{\ \beta\delta}- \tilde{\Gamma}^\delta_{\ \beta\sigma}\tilde{\Gamma}^\rho_{\ \alpha\delta}$.
% \nonumber . 
%\end{eqnarray}
}
which describes the notion of curvature in $\mathcal{U}_4$ completely.
In a similar manner to what was done in the previous Section \ref{Sec:2:A}, one can define the \textit{non-null torsion Ricci tensor} as $\tilde{R}_{\alpha\beta}:=\tilde{R}_{\alpha\sigma\beta}^{\ \ \ \ \sigma}$ as well as the \textit{non-null torsion Ricci scalar} as $\tilde{R}:=\tilde{R}_\alpha^{\ \alpha}$. We note that it is possible to relate the torsion-free Ricci tensor to the non-null torsion Ricci tensor through the use of the Levi-Civita connection and the contorsion tensor. That is, the Ricci tensors can be related through the following expression
\begin{align}\label{rictric2}
R_{\alpha\sigma} = \tilde{R}_{\alpha\sigma} + D_\alpha K^\beta_{\ \beta\sigma} - D_\beta K^\beta_{\ \alpha\sigma}+ K^\rho_{\ \beta\sigma}K^\beta_{\ \alpha\rho} - K^\rho_{\ \alpha\sigma}K^\beta_{\ \beta\rho} \ . 
\end{align}
Lastly, by making use of the above expression, it is possible to relate the torsion-free Ricci scalar, $R$, to the non-null torsion Ricci scalar, $\tilde{R}$. That is, by taking the trace of equation (\ref{rictric2}), we obtain the following expression that relates the two Ricci scalars
\begin{align}\label{ricciscalarec}
R = \tilde{R} - 2D_\alpha T^{\alpha} - T \ , 
\end{align}
where $T^\alpha:=T^{\beta\alpha}_{\ \ \beta}$ is the trace of the torsion tensor and $T$ is the \textit{torsion scalar} which is defined as \cite{arcos}
\begin{align}\label{torsionscalarec}
T := \frac{1}{4}T_\sigma^{\ \beta\alpha}T^\sigma_{\ \beta\alpha} - \frac{1}{2}T^{\beta\alpha}_{\ \ \sigma}T^\sigma_{\ \beta\alpha} - T^{\beta\alpha}_{\ \ \beta}T^\sigma_{\ \alpha\sigma} \ . 
\end{align}
As we will see later on, the torsion scalar defined above plays an important role at the level of the gravitational action when studying specific gravitational theories with non-vanishing torsion \cite{maluf}.
We now turn our attention back to the torsion tensor which is defined in equation (\ref{tordef}). It is common to decompose the torsion tensor into a trace component, a totally antisymmetric component and a third component, $Q_{\alpha\beta\sigma}$, which is antisymmetric in its second and third indices as well as traceless, i.e., $Q^\alpha_{\ \beta\alpha}=0$ \cite{shapiro}. The reason for this is to determine what roles each of the components in the decomposition play in an equation where the torsion tensor appears. For example, in the Dirac equation in Riemann-Cartan space-time, only the totally antisymmetric part of the torsion tensor couples to the fermionic contribution while the trace part and the tensor $Q_{\alpha\beta\sigma}$ decouple completely \cite{shapiro}. The decomposition of the torsion tensor is as follows
\begin{align}\label{decomp1}
T_{\alpha\beta\sigma} = \frac{1}{3}\left(T^\rho_{\ \beta\rho}g_{\alpha\sigma} - T^\rho_{\ \sigma\rho}g_{\alpha\beta}\right) + T_{[\alpha\beta\sigma]} + Q_{\alpha\beta\sigma} \ . 
\end{align}
In the above decomposition, the trace component is $\frac{1}{3}\left(T^\rho_{\ \beta\rho}g_{\alpha\sigma} - T^\rho_{\ \sigma\rho}g_{\alpha\beta}\right)$ and the totally antisymmetric component is $T_{[\alpha\beta\sigma]}$. The tensor field $Q_{\alpha\beta\sigma}$ is defined through the difference.

\subsection{Geodesics in Riemannian space-time}
\label{Sec:2:C}
In this Section, we revise the notion of geodesics in the Riemannian space-time $\mathcal{V}_4$. To begin with, we first discuss the notion of auto-parallels. We say that a curve with tangent vector field $\bm{\xi}$ is an auto-parallel of the Levi-Civita connection provided \cite{hehl76}
\begin{align}\label{gtf1}
D_{\bm{\xi}}\bm{\xi} = 0 \ \ \ \iff \ \ \ \xi^\beta D_\beta\xi^\alpha = 0 \ . 
\end{align}
In addition, we define an extremal curve to be a curve that is such that we have an extremum of the length with respect to the metric tensor, $\bm{g}$ \cite{hehl76}. In our Riemannian space-time, $\mathcal{V}_4$, the tangent vector field of the extremal curve satisfies the well-known equation
\begin{align}\label{gtf2}
\frac{\textup{d}\xi^\alpha}{\textup{d}\lambda} + \Gamma^\alpha_{\ \beta\rho}\xi^\beta\xi^\rho = 0 \ , 
\end{align}
where $\lambda$ is the affine parameter associated with the connection $D$ and the Christoffel symbols $\Gamma^\rho_{\ \alpha\beta}$ are given by equation (\ref{lccs}) \cite{wald4}. It is not difficult to see from equation (\ref{tfcs}) that equations (\ref{gtf1}) and (\ref{gtf2}) are equivalent. Since these two notions coincide in $\mathcal{V}_4$, we say that a curve with tangent vector field $\bm{\xi}$ is a \textit{geodesic} of the Levi-Civita connection $D$ provided that equation (\ref{gtf1}) is satisfied \cite{wald4}. That is, a curve is a geodesic if its tangent vector is parallel propagated in the direction along the curve itself \cite{wald4}. \\ \\

\subsection{Auto-parallels and extremal curves in Riemann-Cartan space-time}
\label{Sec:2:D}
In the previous Section, we discussed how the notions of auto-parallels and extremal curves given the Levi-Civita connection in a Riemannian space-time coincide. Nonetheless, in the case of Riemann-Cartan space-time, these two notions only coincide when the torsion tensor is totally antisymmetric \cite{hehl76}. This fact makes it necessary to distinguish between these two notions in a Riemann-Cartan space-time. On the one hand, we say that a curve with tangent vector field $\bm{\xi}$ is an auto-parallel of the Cartan connection $\nabla$ provided
\begin{align}
\nabla_{\bm{\xi}}\bm{\xi}\ \ \ \iff\ \ \ \xi^\beta\nabla_\beta\xi^\alpha = 0 \ . 
\end{align}
On the other hand, we define extremal curves in the same way as before: a curve with tangent vector field $\bm{\xi}$ is an extremal curve if the equations of motion satisfied by $\bm{\xi}$ yield an extremum of the length with respect to the metric tensor, $\bm{g}$. In gravitational theories constructed on the Weitzenb\"ock space-time, such as $f(T)$ gravity, the paths followed by test particles in the presence of gravity are auto-parallels of the Levi-Civita connection \cite{deandrade,deandrade2}. Accordingly, for such theories
%. Equivalently That is, in curvatureless theories of gravity, 
the equations of motion for free-falling test particles are given by equation (\ref{gtf2}). %For now, however, we do not wish to impose any further conditions on our space-time $\mathcal{U}_4$. 

In the following section, we shall obtain the non-null torsion Raychaudhuri equation following \cite{kar,Capozziello:2001mq} as well as the geometric Raychaudhuri equation following \cite{Dey:2017fld,Luz:2017ldh}. Therein, we shall not assume that the curves are necessarily auto-parallels of the affine connection $\nabla$. Instead, we shall first assume that $\nabla_{\bm{\xi}}\bm{\xi}$ is non-vanishing and derive these Raychaudhuri equations with this assumption in mind.

%%%%%%%%%%%%%%%%%%%%%%%%%%%%%%%%%%%%%%%%%%%%%%

\section{The Raychaudhuri equation in Riemann-Cartan space-time}
\label{Sec:3}

The Raychaudhuri equation in Riemannian space-time, which we refer to here as the torsion-free Raychaudhuri equation, is well-known \cite{wald4,raychaudhuri,ellis}. For its derivation in Riemannian space-time, one considers a smooth congruence of timelike (or null) curves given the Levi-Civita connection $D$. Some remarks about such an instrumental equation are provided in the Appendix \ref{Appendix}.
Although the Levi-Civita connection is not the natural choice of an affine connection in Riemann-Cartan space-times,  the torsion-free Raychaudhuri equation is, in fact, valid in such space-times.
The natural choice of an affine connection in Riemann-Cartan space-time is the Cartan connection, $\nabla$, whose connection coefficients are given in equation (\ref{nntcs}). In what follows, we wish to consider a smooth congruence of timelike curves generated by $\nabla$ in the Riemann-Cartan space-time $\mathcal{U}_4$ defined previously. Therefore, we wish to make use of the Cartan connection when constructing a Raychaudhuri equation for such a congruence. Here, we shall discuss two such Raychaudhuri equations that exist in the literature \cite{Luz:2017ldh,Dey:2017fld,Capozziello:2001mq}. The first, which is derived in \cite{Capozziello:2001mq,kar}, shall be referred to as the non-null torsion Raychaudhuri equation. The second, which is derived in \cite{Luz:2017ldh,Dey:2017fld}, shall be referred to as the geometric Raychaudhuri equation.
\subsection{Non-null torsion Raychaudhuri equation}
Here, we shall make use of the Riemann-Cartan space-time $\mathcal{U}_4$ defined previously. We begin by considering a smooth congruence of timelike curves generated by the Cartan connection, $\nabla$, with tangent vector field $\bm{\xi}$. Here, $\bm{\xi}$ represents a four-velocity and is normalised to unit length with $\bm{g}\left(\bm{\xi},\bm{\xi}\right)=-1$. By making use of the affine connection $\nabla$, we introduce the tensor field $\bm{B}$ of type $(0,2)$ defined in the following way \cite{Capozziello:2001mq}
\begin{align}\label{re1}
B_{\alpha\beta} := \nabla_\beta\xi_\alpha \ . 
\end{align}
By making use of the fact that the tangent vector field is normalised to unit length, it is not difficult to see from the above definition of $\bm{B}$ that we have $\xi^\beta B_{\beta\alpha}=0$. In order to obtain the non-null torsion Raychaudhuri equation, it is necessary to introduce the so-called kinematic quantities. We do this by first introducing the spatial metric, $h_{\alpha\beta}$, defined as 
\begin{align}\label{re6}
h_{\alpha\beta} := g_{\alpha\beta} + \xi_\alpha\xi_\beta \ , 
\end{align}
where $\bm{g}$ is the metric tensor. We now define the \textit{non-null torsion expansion} as
\begin{align}\label{re7}
\theta := h_\alpha^{\ \beta}\nabla_\beta\xi^\alpha \ , 
\end{align}
the \textit{non-null torsion shear} as
\begin{align}\label{re8}
\sigma_{\alpha\beta} := h^\sigma_{\ \alpha}h^\rho_{\ \beta}\nabla_{(\sigma}\xi_{\rho)} - \frac{\theta}{3}h_{\alpha\beta} \ , 
\end{align}
and the \textit{non-null torsion twist} as
\begin{align}\label{re9}
\omega_{\alpha\beta} := h^\sigma_{\ \alpha}h^\rho_{\ \beta}\nabla_{[\sigma}\xi_{\rho]} \ . 
\end{align}
The expansion, shear and twist defined above are referred to here as the \textit{non-null torsion kinematic quantities} and are used to construct the non-null torsion Raychaudhuri equation. In terms of these kinematic quantities, we can write the tensor field $\bm B$ as follows
\begin{align}\label{re22}
B_{\alpha\beta} = \omega_{\alpha\beta} + \sigma_{\alpha\beta} + \frac{\theta}{3}h_{\alpha\beta} - \xi^\rho\xi_\beta B_{\alpha\rho} \ . 
\end{align}
Equation (\ref{re22}) gives the desired decomposition of $B_{\alpha\beta}$ in terms of the non-null torsion expansion, shear and twist. Through the use of equation (\ref{re22}), one can obtain an expression for $\nabla_{\bm\xi}B_{\alpha\beta}$ in terms of the non-null torsion kinematic quantities. One can then take the trace of the resulting expression in order to arrive at the \textit{non-null torsion Raychaudhuri equation}
\begin{align}\label{re42}
\frac{\textup{d}\theta}{\textup{d}\lambda} &= \nabla_\alpha\left(\xi^\beta\nabla_\beta\xi^\alpha\right) - \sigma_{\alpha\beta}\sigma^{\alpha\beta} + \omega_{\alpha\beta}\omega^{\alpha\beta}- \frac{1}{3}\theta^2- \tilde{R}_{\alpha\beta}\xi^\alpha\xi^\beta\nonumber\\ &- T^\rho_{\ \beta\alpha}\xi^\beta\left(\sigma^\alpha_{\ \rho} + \omega^\alpha_{\ \rho} + \frac{1}{3}\theta h^\alpha_{\ \rho} - \xi^\sigma\xi_\rho\nabla_\sigma\xi^\alpha\right) \ . 
\end{align}
Equation (\ref{re42}) gives a generalized Raychaudhuri equation for a smooth congruence of timelike curves generated by the Cartan connection, $\nabla$, and has been derived before in \cite{Capozziello:2001mq,kar}.

Equation (\ref{re42}) can be particularised in the special case of a smooth congruence of timelike auto-parallels of the Cartan connection $\nabla$, i.e., $\xi^\beta\nabla_\beta\xi^\alpha=0$. By requiring that $\xi^\alpha$ satisfies this condition, it follows that the Raychaudhuri equation (\ref{re42}) reduces to
\begin{eqnarray}\label{refn}
\frac{\textup{d}\theta}{\textup{d}\lambda}=  - \sigma_{\alpha\beta}\sigma^{\alpha\beta} - \frac{1}{3}\theta^2 - \tilde{R}_{\alpha\beta}\xi^\alpha\xi^\beta- \frac{1}{3}T_\beta\xi^\alpha\left(\delta^\beta_\alpha\theta - \sigma^\beta_{\ \alpha}\right) - Q_{\rho\beta\alpha}\xi^\beta\sigma^{\alpha\rho} \ , 
\end{eqnarray}
where we have made use of the decomposition of the torsion tensor given in equation (\ref{decomp1}). We immediately notice that the totally antisymmetric part of the torsion tensor, $T_{[\alpha\beta\rho]}$, does not make any contribution in the Raychaudhuri equation. This is due to the fact that $T_{[\alpha\beta\rho]}$ is totally antisymmetric and that a summation is carried out with these components together with symmetric components contained in brackets. This causes the contribution involving the totally antisymmetric part of the torsion tensor to vanish. We are, however, left with contributions from the trace part of the torsion tensor as well as with contributions from the tensor field $Q_{\alpha\beta\rho}$.

It is important to note that, while the non-null torsion Raychaudhuri equation is valid in the Riemann-Cartan space-time, the non-null torsion kinematic quantities do not provide a geometric description of the congruence. Therefore, the non-null torsion Raychaudhuri equation does not describe the geometric expansion. In order to study the geometric rate of expansion, we turn our attention to the Raychaudhuri equation derived in \cite{Luz:2017ldh,Dey:2017fld}.
\subsection{Geometric Raychaudhuri equation}
Following the derivation given in \cite{Luz:2017ldh,Dey:2017fld}, we consider the tensor field $\tilde{\bm{B}}$ of type $(0,2)$ defined in the following way
\begin{align}\label{re1}
\tilde B_{\alpha\beta} := \nabla_\beta\xi_\alpha+T_{\alpha\nu\beta}\xi^\nu=D_\beta\xi_\alpha+K_{\alpha\nu\beta}\xi^\nu \ . 
\end{align}
Given the fact that the tangent vector field is normalised to unit length, one can show that $\xi^\alpha \tilde B_{\alpha\beta}=T_{(\alpha\nu)\beta}\xi^\alpha\xi^\nu$. Furthermore, due to the antisymmetric property of the torsion tensor, we have $\xi^\beta \tilde B_{\alpha\beta}=\xi^\beta\nabla_\beta\xi_\alpha$ which vanishes when considering auto-parallel curves of the Cartan connection. We now wish to find a physical interpretation of $\tilde{\bm B}$. By denoting the deviation vector field as $\bm \eta$, we can make use of the Lie derivative in order to obtain
\begin{align}\label{lie_use}
\pounds_\xi\eta^\beta=\xi^\alpha\nabla_\alpha\eta^\beta-K^\beta_{\ \alpha\nu}\eta^\nu\xi^\alpha-\eta^\alpha\nabla_\alpha\xi^\beta+K^\beta_{\ \alpha\nu}\eta^\alpha\xi^\nu=0\ .
\end{align}
It follows from equations (\ref{lie_use}) and (\ref{re1}) that one can write
\begin{align}
\xi^\alpha\nabla_\alpha\eta^\beta=\left(\nabla_\alpha\xi^\beta+T^\beta_{\ \nu\alpha}\xi^\nu\right)\eta^\alpha=\tilde B^\beta_{\ \alpha}\eta^\alpha\ .
\end{align}
From the last expression, one can interpret the tensor field $\tilde{\bm B}$ as being a measure of the failure for the deviation vector field $\bm\eta$ to be parallel propagated with respect to the Cartan connection.

We now define the \textit{geometric expansion} as
\begin{align}\label{re7L}
\tilde\theta := h_\alpha^{\ \beta}\tilde B^\alpha_{\ \beta} \ , 
\end{align}
the \textit{geometric shear} as
\begin{align}\label{re8L}
\tilde\sigma_{\alpha\beta} := h^\sigma_{\ \alpha}h^\rho_{\ \beta}\tilde B_{(\sigma\rho)} - \frac{\theta}{3}h_{\alpha\beta} \ , 
\end{align}
and the \textit{geometric twist} as
\begin{align}\label{re9L}
\tilde\omega_{\alpha\beta} := h^\sigma_{\ \alpha}h^\rho_{\ \beta}\tilde B_{[\sigma\rho]} \ . 
\end{align}
These physical quantities, referred to here as the \textit{kinematic quantities}, allow us to write the tensor field $\tilde{\bm B}$ as follows
\begin{align}\label{re22L}
\tilde B_{\alpha\beta} = \tilde\omega_{\alpha\beta} + \tilde\sigma_{\alpha\beta} + \frac{\tilde\theta}{3}h_{\alpha\beta} - \xi^\rho\xi_\beta \tilde B_{\alpha\rho}- \xi^\rho\xi_\alpha \tilde B_{\rho\beta} \ . 
\end{align}
By making use of equation (\ref{re22L}), one can find an expression for the type $(0,2)$ tensor field $\nabla_{\bm\xi}\tilde{B}_{\alpha\beta}$. One can then take the trace of the resulting expression and obtain the \textit{geometric Raychaudhuri equation} given below
\begin{align}\label{re42L}
\frac{\textup{d}\tilde\theta}{\textup{d}\lambda} &= \nabla_\alpha\left(\xi^\beta\nabla_\beta\xi^\alpha\right) - \tilde\sigma_{\alpha\beta}\tilde\sigma\phantom{}^{\alpha\beta} + \tilde\omega_{\alpha\beta}\tilde\omega\phantom{}^{\alpha\beta}- \frac{1}{3}\tilde\theta\phantom{}^2- \tilde{R}_{\alpha\beta}\xi^\alpha\xi^\beta\nonumber\\ &-T_{\alpha\beta\rho}\xi^\beta\left(\frac13\tilde\theta h^{\alpha\rho}+\tilde\sigma\phantom{}^{\alpha\rho}-\tilde\omega\phantom{}^{\alpha\rho}-\xi^\alpha\xi^\nu\nabla_\nu\xi^\rho\right)+\xi^\alpha\nabla_\alpha\left(T_\nu\xi^\nu\right)\ .
\end{align}
Equation (\ref{re42L}) was derived in \cite{Luz:2017ldh,Dey:2017fld} and gives us a generalized notion of the expansion scalar $\tilde\theta$ rate when considering Riemann-Cartan space-time.

Up until now, we have not given a physical interpretation of what is described by the geometric expansion scalar, shear tensor or twist tensor. In order to give such a physical interpretation, we consider a cross-sectional area, $\mathcal{A}$, that encloses our congruence of timelike curves. In addition, consider an observer moving along a curve contained in our smooth congruence. The expansion scalar, $\tilde\theta$, provides us with a description as to how the cross-sectional area, $\mathcal{A}$, expands as well as contracts as observed by our observer moving along a curve contained in the congruence. The shear tensor, $\tilde\sigma_{\alpha\beta}$, on the other hand provides us with a description of the tendency for $\mathcal{A}$ to become ellipsoidal in shape as seen by our observer. Finally, the twist tensor, $\tilde\omega_{\alpha\beta}$, describes how the cross-sectional area, $\mathcal{A}$, twists about the congruence it encloses as seen by our observer.

Having derived the non-null torsion Raychaudhuri equation (\ref{re42}) as well as the geometric Raychaudhuri equation (\ref{re42L}), it is trivial to obtain the torsion-free Raychaudhuri equation\footnote{The reader seeking a derivation of the torsion-free Raychaudhuri equation in the context of Riemannian space-time is directed to \cite{wald4}. In order to derive the torsion-free Raychaudhuri equation, we can simply consider what happens to either equation (\ref{re42}) or equation (\ref{re42L}) in the absence of torsion since the Cartan connection $\nabla$ is nothing more than the Levi-Civita connection, $D$, when we have a vanishing torsion tensor.}. At this stage, it is important to note that, since the Levi-Civita connection may be thought of as an additional structure in the Riemann-Cartan space-time $\mathcal{U}_4$, the torsion-free Raychaudhuri equation (\ref{re42lc}) - see Appendix \ref{Appendix} for its full derivation-  may be useful when considering non-null torsion theories of gravity. On the one hand, when considering a smooth congruence of timelike curves generated by $\nabla$, we make use of equation (\ref{re42L}) to describe the expansion rate of a cross-sectional area about the congruence. On the other hand, when considering a smooth congruence of timelike curves generated by $D$, we make use of equation (\ref{re42lc}) to describe the expansion rate of a cross-sectional area about the congruence. Since we are interested in considering test particles when studying the focusing conditions, we will make use of the torsion-free Raychaudhuri equation in our analysis since test particles are taken to follow extremal curves which are auto-parallels of the Levi-Civita connection.
\section{$f(T)$ theories of gravity}
\label{Sec:4}

As explained in the Introduction, $f(T)$ theories of gravity are constructed by making use of the Weitzenb\"ock space-time, where the non-null torsion Riemann tensor vanishes and the torsion tensor is non-vanishing.  In Section \ref{Sec:4:A} we discuss the Weitzenb\"ock connection and in Section \ref{Sec:4:B} we discuss the formulation of the $f(T)$ theories of gravity. In Section \ref{Sec:5}, we consider the Robertson-Walker space-time.

% We therefore wish to begin our discussion of $f(T)$ theories of gravity by introducing the Weitzenb\"ock connection. 

\subsection{Tetrad formalism and the Weitzenb\"ock connection}
\label{Sec:4:A}

In order to construct the Weitzenb\"ock connection, we introduce a nonholonomic basis of smooth vector fields, $e^a_{\ \sigma}$, that are such that 
\begin{align}\label{tetd}
g_{\mu\nu} = \eta_{ab}e^a_{\ \mu}e^b_{\ \nu} \ , 
\end{align}
where % $g_{\mu\nu}$ is the metric tensor and 
$\eta_{ab}=\textup{diag}(-1,1,1,1)$ is the Minkowski metric and $\{e^a_{\ \sigma}\}$ is referred to as a tetrad \cite{wald4}. Given a metric tensor, $g_{\mu\nu}$, we can construct tetrads according to equation (\ref{tetd}). Nonetheless, it is important to note that there does not necessarily exist a unique tetrad construction for a given metric tensor. Moreover, in general, tetrads $e^a_{\ \sigma}$ are not invariant under Lorentz transformations \cite{tamanini}. In the realm of $f(T)$ theories of gravity, quantities are derived from tetrads rather than from the metric tensor and, therefore, since tetrads are not in general Lorentz invariant, these quantities may not necessarily be Lorentz invariant. This implies that the field equations in $f(T)$ gravity are not necessarily Lorentz invariant ({\it c.f.} \cite{tamanini} for a detailed discussion on this issue). For now, however, we can discuss the notion of curvatureless gravity, i.e., gravitational theories with vanishing curvature, by constructing a connection that yields a vanishing non-null torsion Riemann tensor through the use of tetrads. 
By considering the Riemann-Cartan space-time $\mathcal{U}_4$ and imposing that the connection coefficients, $\tilde{\Gamma}^\rho_{\ \mu\nu}$, of the affine connection, $\nabla$, to be of the form
\begin{align}\label{weitz}
\tilde{\Gamma}^\rho_{\ \mu\nu} = e_a^{\ \rho}\partial_\mu e^a_{\ \nu} \,,
\end{align}
one guarantees that the non-null torsion curvature $\tilde{R}_{\alpha\beta\nu}^{\ \ \ \ \mu}$ as defined in equation (\ref{nntrd}) vanishes \cite{weitz}. 
Definition (\ref{weitz}) is referred to as the \textit{Weitzenb\"ock connection}. 
Following the definition given in \cite{tseng} we refer to the tuple $\left(\bm{\mathcal{M}}',\bm{g},\nabla\right)$, where $\nabla$ is now the Weitzenb\"ock connection, as the \textit{Weitzenb\"ock space-time} and we denote this space-time as $\mathcal{W}_4$.
%
%
%From equation (\ref{rictric2}), we can write the torsion-free Ricci tensor, $R_{\mu\rho}$, as \cite{uli} 
%\begin{align}{\label{tele7}}
%R_{\mu\rho}&= D_\mu K^\nu_{\ \nu\rho} - D_\nu K^\nu_{\ \mu\rho}  + K^\alpha_{\ \nu\rho}K^\nu_{\ \mu\alpha}\nonumber\\ &- K^\alpha_{\ \mu\rho}K^\nu_{\ \nu\alpha} \ . 
%\end{align}
%Furthermore, 

Consequently, it is easy to conclude that equation (\ref{ricciscalarec}) in $\mathcal{W}_4$ renders the following expression for the torsion-free Ricci scalar
\begin{align}\label{rsctsc}
R = -2D_\mu T^{\nu\mu}_{\ \ \nu} - T \ , 
\end{align}
where $T$ is the torsion scalar defined in equation (\ref{torsionscalarec}). By defining the Weitzenb\"ock contorsion tensor, $W^{\rho}_{\ \mu\nu}$, as
\begin{eqnarray}
W_{\rho\mu\nu} := \frac{1}{2}(-T_{\rho\mu\nu} - T_{\mu\nu\rho} + T_{\nu\rho\mu}) \label{weitzk} = {K}_{\rho\mu\nu} - {T}_{\rho\mu\nu} \ , \label{weitzk1} 
\end{eqnarray}
we can introduce the tensor
\begin{align}\label{stensor}
S_\rho^{\ \mu\nu} := \frac{1}{2}(W^{\mu\nu}_{\ \ \ \rho} + \delta^\mu_\rho T^{\beta\nu}_{\ \ \ \beta} - \delta^\nu_\rho T^{\beta\mu}_{\ \ \ \beta} ) \ , 
\end{align}
known as the \textit{superpotential} in order to write the torsion scalar $T$ as \cite{arcos,maluf}
\begin{align}
T = T^\rho_{\ \mu\nu}S_\rho^{\ \mu\nu} \ . 
\end{align}
We are now in a position to discuss the formulation of $f(T)$ theories of gravity which is done in the following Section. 

\subsection{Lagrangian formulation}
\label{Sec:4:B}
We begin this discussion by first considering the Einstein-Hilbert (EH) action in GR. We will then make use of this action to show that GR is, in fact, equivalent to Teleparallel Gravity, $f(T)=T$, following \cite{cai2015}. The Einstein-Hilbert action, $S_{EH}$, is given by the following expression 
\begin{align}\label{ehaction}
S_{\rm EH} = \frac{1}{16\pi}\int\textup{d}^4x\ R\ \sqrt{-g} \ , 
\end{align}
where $R$ is the torsion-free Ricci scalar and $g$ is the determinant of the metric tensor \cite{hobson}.
From equation (\ref{rsctsc}), which gives an expression for the torsion-free Ricci scalar in terms of torsion terms, the EH action can be written as 
\begin{align}
S_{\rm EH} = -\frac{1}{16\pi}\int\textup{d}^4x\ e\,T - \frac{1}{8\pi}\int\textup{d}^4x\ e\,D_\mu T^\mu \ , 
\end{align}
where $e$ is the determinant of the tetrad field, $e^a_{\ \sigma}$. Now, since the right-hand side second term in the previous expression contains a total derivative, $D_\mu T^\mu$, with respect to the volume element $e\ \textup{d}^4x$ it makes no contribution. Therefore, the EH action reduces to the well known action of Teleparallel Gravity. This result is usually referred to as the equivalence between GR and the Teleparallel Gravity, also dubbed the Teleparallel Equivalent of General Relativity (TEGR) \cite{cai2015}.  We note here that if one were to consider a general function, $f(R)$, of the torsion-free Ricci scalar in equation (\ref{ehaction}), one would obtain the action of so-called $f(R)$ theories of gravity (see \cite{sotiriou} for a review). Similarly, one can consider a general function, $f(T)$, \footnote{Although GR and TEGR are equivalent, a given $f(R)$ theory of gravity is not in general equivalent to the corresponding $f(T)$ theory of gravity since the two expressions may not differ by a total derivative.} of the torsion scalar \cite{cai2015,bengochea} and obtain 
\begin{align}\label{ftac}
S = \int \textup{d}^4x\ e\left(\frac{1}{16\pi}f(T) + \mathcal{L}_m\right) \ , 
\end{align}
which is the action in $f(T)$ theories of gravity with the matter content in the matter Lagrangian $\mathcal{L}_m$. By varying the action given in equation (\ref{ftac}) with respect to the vierbein $e^l_{\ \sigma}$, the $f(T)$ field equations
\begin{eqnarray}\label{fe12}
&\frac{1}{4}f\delta^\sigma_\alpha+ f_T\left[T^\nu_{\ \beta\alpha}S_\nu^{\ \sigma\beta} + e^{-1}e^l_{\ \alpha}\partial_\nu\left(e\,e_l^{\ \beta}S_\beta^{\ \sigma\nu}\right)\right]+ f_{TT}S_\alpha^{\ \sigma\nu}\partial_\nu T= -4\pi\,\frac{e^l_{\ \alpha}}{e}\frac{\delta (e\mathcal{L}_m)}{\delta e^l_{\ \sigma}}=:4\pi\,{\mathcal{T}}^\sigma_\alpha \ ,\nonumber\\
&
\end{eqnarray}
are obtained where $\mathcal{T}_{\alpha}^{\ \sigma}$ is the energy-momentum tensor and 
$f_T$ and $f_{TT}$ are the first and second order derivative of $f(T)$ with respect to $T$ respectively \cite{cai2015}.

Having discussed the formulation of the $f(T)$ theories of gravity, we now turn our attention to studying the focusing conditions in the context of these theories. In carrying out such a study, we shall consider the metric tensor to represent a spatially-flat Robertson-Walker space-time.

%%%%%%%%%%%%%%%%%%%%%%%%%%%%%%%%%%%%%%%%%%%%%%%%%
\subsection{Robertson-Walker space-time in $f(T)$ gravity}
\label{Sec:5}
In order to deal with cosmological scenarios, it makes sense to specify the space-time to be a Robertson-Walker space-time and study $f(T)$ phenomenology in such a structure. Here, we make use of the Robertson-Walker space-time in the case of spatially flat geometry which is given by the following metric \cite{wald4}
\begin{align}
\textup{d}s^2 = -\textup{d}t^2 + a^2(t)\left(\textup{d}x^2 + \textup{d}y^2 + \textup{d}z^2\right) \ . 
\end{align}
By making use of the above metric together with equation (\ref{tetd}), we note that we can make use of the following diagonal tetrad construction
\begin{eqnarray}
e^a_{\ \mu} = {\rm diag}(1,\;a(t),\;a(t),\;a(t) )
%\begin{pmatrix} 
%1 & 0 & 0 & 0 \\
%0 & a(t) & 0 & 0 \\
%0 & 0 & a(t) & 0 \\
%0 & 0 & 0 & a(t)
%\end{pmatrix} \ . 
\end{eqnarray}
Although in the following we shall consider congruences whose tangent vector fields are not necessarily those of a co-moving observer, we can still make use of the diagonal tetrad construction given above since we will always take the matter field to be co-moving. As widely known, for the diagonal tetrad construction given above, the torsion scalar has the following expression \cite{tamanini}
\begin{align}\label{trw}
T = 6H^2 \ ,
\end{align}
where we have defined the Hubble parameter as $H:={\dot{a}}/{a}$ \cite{baumann}. In this note, we consider an energy-momentum tensor that is of the general perfect fluid form \cite{tamanini,baumann}
\begin{align}\label{emtgpfm}
\mathcal{T}_{\mu\nu}=\rho\chi_\mu\chi_\nu+P\left(\chi_\mu\chi_\nu+g_{\mu\nu}\right)\ ,
\end{align}
where $\chi^\mu:=(1,0,0,0)$ is taken to be the four-velocity associated with a co-moving observer. The non-vanishing $f(T)$ field equations components associated with a spatially-flat Robertson-Walker space-time for which an energy-momentum tensor is of the form (\ref{emtgpfm}) become \cite{tamanini}
\begin{align}
\frac{f}{4} - 3H^2f_T &= -4\pi\rho \ , \label{feq1} \\
\frac{f}{4}-3H^2f_T&=\dot{H}\left(f_T+12H^2f_{TT}\right)+4\pi P\ .\label{feq2}
\end{align}
Having obtained the relevant $f(T)$ field equations, in the following Section we turn our attention to the consideration of the focusing conditions.

%%%%%%%%%%%%%%%%%%%%%%%%%%%%%%%%%%%%%%%%%%%%%%%%%
\section{Focusing conditions in $f(T)$ gravity}
\label{Sec:6}

As mentioned in Section \ref{Sec:2:D}, for any $f(T)$ theory the differential equation, referred to as the \textit{force equation} \cite{arcos}, which describes the motion of a test particle in the presence of gravity, coincides with that which describes auto-parallel curves of the Levi-Civita connection \cite{hehl76}. We therefore turn our attention to studying the torsion-free Raychaudhuri equation since the cross-sectional area that it describes encloses a congruence of auto-parallels of the Levi-Civita connection. More specifically, we shall consider equation (\ref{re42lcc}), i.e., the Raychaudhuri equation for a congruence of timelike auto-parallels of the Levi-Civita connection whose tangent vector field, $\bm{\xi}$, is taken to be hypersurface orthogonal. Since it is the torsion-free expansion scalar that provides us with the desired description of the divergence or convergence of neighbouring paths followed by test particles, the condition of attractive gravity would be given by the inequality
\begin{align}
\frac{\textup{d}\overset{\circ}\theta}{\textup{d}\tau}\leq0\ .
\end{align}
Let us now turn our attention to the right-hand side of equation (\ref{re42lcc}). It is evident that the first two terms there are nonpositive, while the sign of the third term is not immediately clear. As widely known, the focusing conditions (or energy conditions in the case of GR) arise as a result of requiring that this third term be nonpositive, i.e., $R_{\mu\nu}\xi^\nu\xi^\nu\geq0$ \cite{wald4}. One possibility that allows for this inequality to hold is to require that the Einstein tensor, $G_{\mu\nu}:=R_{\mu\nu}-\frac{1}{2}Rg_{\mu\nu}$, contracted with the tangent vector field is nonpositive, i.e., $G_{\mu\nu}\xi^\mu\xi^\nu\geq0$ \cite{wald4}. This is referred to as the \textit{weak focusing condition}. Alternatively, the inequality $G_{\mu\nu}\xi^\mu\xi^\nu\geq-\frac{1}{2}G$ also allows for the third term on the right-hand side of equation (\ref{re42lcc}) to be nonpositive. This inequality is referred to as the \textit{strong focusing condition}. 

In this note, we are interested in studying the satisfaction/violation of the weak and strong focusing conditions for one-parameter dependent congruences of timelike auto-parallels of the Levi-Civita connection whose tangent vector field is given by the expression \cite{alvaro2}
\begin{align}\label{onepara}
\xi^\alpha=\gamma\left(1,\beta,0,0\right)\ , \; {\rm with}\;\; \gamma:=\frac{1}{\sqrt{1-a^2\beta^2}}\ .
\end{align}
We require that the tangent vector field $\xi^\alpha$ be parallel propagated with respect to the Levi-Civita connection, i.e., $\xi^\alpha D_\alpha\xi^\nu=0$. Such a requirement implies that the parameter $\beta$ is of the form
\begin{align}
\beta=\frac{C}{a\sqrt{a^2+C^2}}\ ,
\label{beta}
\end{align}
where $C\in\mathbb{R}$ is the parameter that, when varied, produces different tangent vector fields that are parallel propagated with respect to the Levi-Civita connection. We note that when $C=0$ the four-velocity $\xi^\alpha$ given in equation (\ref{onepara}) reduces to the four-velocity associated with a co-moving observer, $\chi^\alpha$, used in the construction of the energy-momentum tensor given in equation (\ref{emtgpfm}). The congruence whose tangent vector field corresponds to a co-moving observer is often referred to as a \textit{fundamental congruence}. The $f(T)$ focusing conditions for the case of a fundamental congruence, i.e., when $C=0$, have been derived before in \cite{liu}. In this note, we shall derive both the weak and strong focusing conditions for a general value of the parameter $C$.

{\bf Weak focusing condition} We first consider the inequality $G_{\mu\nu}\xi^\mu\xi^\nu\geq0$. By making use of equation (\ref{rictric2}) for the case where $\tilde{R}_{\mu\nu}=0$, one can obtain the following expression for the Einstein tensor
\begin{align}\label{EinsteinTensorWeitzenbock}
G_{\mu\nu}&=D_\nu T_\mu-D_\mu T_\nu+2D_\beta S_{\nu\mu}^{\ \ \beta}-D_\beta T^\beta_{\ \mu\nu}+\frac{1}{2}Tg_{\mu\nu}+K^\alpha_{\ \beta\nu}K^\beta_{\ \mu\alpha}+K^\alpha_{\ \mu\nu}T_\alpha\ .
\end{align}
The above expression is a completely geometric statement for the Weitzenb\"ock space-time $\mathcal{W}_4$. Therefore 
(\ref{EinsteinTensorWeitzenbock}) does not depend upon the specific $f(T)$ model constructed on $\mathcal{W}_4$. Nonetheless, at this stage let us introduce  in  (\ref{EinsteinTensorWeitzenbock})  the energy-momentum tensor by substituting equation (\ref{fe12}), which are the $f(T)$ field equations, into the above expression. This yields 
% the following expression for the Einstein tensor which is specific to $f(T)$ theories of gravity
%
%
\begin{align}
\label{etenft}
G_{\mu\nu}&=\frac{1}{f_T}\left(k^2\tau_{\mu\nu}-2f_{TT}S_{\nu\mu}^{\ \ \beta}\partial_\beta T-\frac{1}{2}fg_{\mu\nu}\right)+W^\beta_{\ \lambda\nu}S_{\beta\mu}^{\ \ \lambda}+D_\nu T_\mu-D_\mu T_\nu-D_\lambda T^\lambda_{\ \mu\nu}\nonumber\\
&+\frac{1}{2}Tg_{\mu\nu}+K^\alpha_{\ \beta\nu}K^\beta_{\ \mu\alpha}+K^\alpha_{\ \mu\nu}T_\alpha\ .
\end{align}
By contracting the above expression with the four-velocity given in equation (\ref{onepara}), the weak focusing condition holds if and only if we have
\begin{align}\label{weakonepara}
8\pi\rho_{eff}:=\frac{1}{f_T}\bigg[8\pi\rho+\frac{1}{2}\left(f-f_TT\right)+a^2\beta^2\bigg(2Hf_{TT}\partial_t T-\frac{1}{2}f\bigg)\bigg]+3a^2\beta^2H^2\geq0\ ,
\end{align}
where we have introduced the so-called \textit{effective energy density}, $\rho_{eff}$\footnote{As expected, in the case of TEGR, i.e., $f(T)=T$, the weak focusing condition given above reduces to the usual weak energy condition, $\rho\geq0$.}. 

{\bf Strong focusing condition} Equipped with the discussion above, we now study the inequality $G_{\mu\nu}\xi^\mu\xi^\nu\geq-\frac{1}{2}G$. It is not difficult to show that the strong focusing condition holds if and only if we have
\begin{align}\label{strongonepara}
4\pi\left(\rho_{eff}+3P_{eff}\right)&:=\frac{1}{f_T}\bigg[4\pi\left(\rho+3P\right)-\frac{1}{2}\left(f-f_TT\right)+3f_{TT}H\partial_t T\nonumber\\ &+a^2\beta^2\Big(2Hf_{TT}\partial_t T-\frac{1}{2}f\Big)\bigg]+3a^2\beta^2H^2\geq0\ ,
\end{align}
where we have introduced the so-called \textit{effective pressure}, $P_{eff}$\footnote{As expected, the strong focusing condition reduces to the strong energy condition $\rho+3P\geq0$ for the case of TEGR.}.  It is important to note that the effective energy density and effective pressure contain, in addition to the energy-matter content, terms that depend on the geometry. Therefore, one cannot impose the same physical conditions on the effective energy density and effective pressure that one would impose on the energy density $\rho$ and pressure $P$. For a discussion on the effective energy-momentum tensor and energy conditions in modified theories of gravity, the reader is directed to \cite{Capozziello:2013vna,Capozziello:2014bqa} 

Before examining these two focusing conditions for three $f(T)$ paradigmatic cosmological models, a pertinent remark would be to stress that in both 
(\ref{weakonepara}) and (\ref{strongonepara}) the dependence on the one-parameter $C$, which label the timelike auto-parallels of the Levi-Civita connection, appears through the parameter $\beta$ defined in (\ref{beta}). For the case of a fundamental congruence ($C=0$), one has $\beta=0$ and then arrives at the results given in \cite{liu}

%%%%%%%%%%%%%%%%%%%%%%%%%%%%%%%%%%%%%%%%%%%%%%%%%
\section{$f(T)$ cosmological models}
\label{Sec:7}

In this Section, we study the focusing conditions for three specific bi-parametric $f(T)$ cosmological models.

Here, we will allow the $f(T)$ function describing a specific cosmological model to depend on two values: $\sigma$ and $\alpha$. By changing the values of $\alpha$ and $\sigma$ in the $f(T)$ function, one obtains different $f(T)$ theories. 

For each choice of cosmological model considered, we first solve the field equation (\ref{feq2}) for the dimensionless Hubble parameter defined as
\begin{align}
\label{hz}
h(z):=\frac{H(z)}{H_0}\ , 
\end{align}
where $z$ denotes the cosmological redshift  $z=a^{-1}-1$ and $H_0$ is the present-day Hubble parameter value \cite{planck, baumann}. 
For a given cosmological model, we shall obtain a solution $h(z)$ for each combination of $\alpha$ and $\sigma$ values. 
In this note, we are interested in studying the focusing conditions for the case of dust, i.e., the fluid in equations (\ref{feq1}) and (\ref{feq2}) would be pressureless, so the 
%
%we have vanishing pressure. Since we are considering the case of vanishing pressure, the conservation equation $D_\nu\mathcal{T}^\nu_{\ \mu}=0$ implies that
% 
the energy density takes the form $\rho=\rho_0\,a^{-3}$ where $\rho_0$ is a constant \cite{baumann}. 
Let us now turn our attention to the field equation (\ref{feq2}) in order to obtain the dimensionless Hubble parameter $h(z)$. We are particularly interested in studying the focusing conditions at the present day value of $z=0$. 
%
% Here, we use Mathematica to solve the differential equation (\ref{feq2}) numerically. 
%
In solving such a differential equation, we impose the initial condition: $h(1100)=h_\Lambda(1100)$, where
\begin{align}
h_\Lambda(z)=\sqrt{{^\Lambda\!\Omega}_{m_0}\left(1+z\right)^3+1-{^\Lambda\!\Omega}_{m_0}}\ ,
\end{align}
is the $\Lambda$CDM solution \cite{baumann}. In addition, for illustrative purposes we take the $\Lambda$CDM energy density parameter to be ${^\Lambda\!\Omega}_{m_0}=0.315$ as found in \cite{planck}. Once (\ref{feq2}) has been solved in order to obtain $h(z)$ values for a particular cosmological model, we can then make use of the field equation (\ref{feq1}) in order to find the energy denstiy parameter values $\Omega_{m_0}:=\frac{8\pi\rho_0}{3H_0^2}$ \cite{baumann}. Since we expect to obtain different $\Omega_{m_0}$ values for different $f(T)$ functions, the parameter $\Omega_{m_0}$ will depend on the values of $\alpha$ and $\sigma$ as described by equation (\ref{feq1}). Model-dependent explicit expressions for  $\Omega_{m_0}$ will be provided in the following.

In this note, we shall consider three paradigmatic $f(T)$ cosmological models with the first being of a polynomial form \cite{yang22}, the second being of an exponential form \cite{bamba} and the third being of a hyperbolic tangent form \cite{wu2010}. In constructing these cosmological models, we make use of the dimensionless torsion scalar, $\bar{T}$, defined through the expression
\begin{align}
\bar{T}:=\frac{T}{6H_0^2}\ .
\end{align}
Before proceeding with the aforementioned $f(T)$ cosmological models, we shall impose two theoretical constraints on the models under consideration.
These constraints are referred to as the \textit{viability conditions}
and have been discussed in the context of $f(R)$ theories of gravity in \cite{albafnd,sotiriou,dolgov2003}. In this note, we make use of the $f(R)$ viability conditions\footnote{
In the context $f(R)$ theories of gravity, an additional viability conditions states that $f_{RR}$ cannot be negative in order to ensure a stable model against the so-called Dolgov-Kawasaki instability \cite{dolgov2003,sotiriou}. 
%This is based on studying the trace of the $f(R)$ field equations, which contains second order derivatives, and identifying the square of an effective mass term whose sign depends on that of $f_{RR}$ \cite{sotiriou}. This leads to the requirement that $f_{RR}$ be nonnegative in order to ensure a stable theory \cite{sotiriou}. 
Nonetheless, such a condition does not have a counterpart  on the sign of $f_{TT}$
.}, as given in \cite{sotiriou}, in order to impose viability conditions on $f(T)$ theories. The first viability condition to impose on an $f(T)$ theory reads
\begin{align}\label{viability1}
\lim_{T\rightarrow\infty}\frac{f(T)}{T}=1 \ ,
\end{align}
and ensures that the $f(T)$ theory of gravity behaves like the TEGR theory of gravity in the limit as $T\rightarrow\infty$, i.e., at early stages of the cosmological evolution.
Complementarily, the second viability condition to be imposed reads \cite{wu2010}
\begin{align}\label{viability2}
f_T > 0 \ .
\end{align}
As can be seen in  equations (\ref{weakonepara}) and  (\ref{strongonepara}), 
this condition ensures that the contribution of the energy matter content in the focusing conditions is always nonnegative, i.e., the {\it effective Newton's constant} $G_{eff}:=G_N/f_T$ is always positive.

Finally, recent developments in the field of Gravitational Waves have placed strong constraints on the gravitational wave speed \cite{ezquiaga}. It is shown in \cite{chunlong} that such a constraint is trivially satisfied for $f(T)$ gravities.
Consequently, the cosmological models discussed below are constructed in such a way that the viability conditions given in equations (\ref{viability1}) and (\ref{viability2}) are satisfied. In addition, in the limit where a cosmological model behaves as TEGR plus a cosmological constant, the parameter $\sigma$ will take on the role of the cosmological constant\footnote{Due to the imposed initial condition $h(z)=h_\Lambda(z)$, $\sigma$ values need to be nonnegative as seen in the following.}.

%%%%%
\subsection{Model 1: $f=6H_0^2\left(\bar{T}+\sigma\bar{T}^\alpha\right)$}
\label{Sec:7:A}
The first cosmological model that we wish to consider is of the polynomial form presented in \cite{linder,yang22}. Herein we restrict our consideration of $\alpha$ and $\sigma$ values to the region $(\sigma,\alpha)\in[0,1]\times[-1,1)$. We note that the viability conditions (\ref{viability1}) and (\ref{viability2}) for the combinations of $\alpha$ and $\sigma$ values considered are satisfied. We also note that the viability conditions are satisfied for larger values of $\sigma$, however, the analysis regarding these larger $\sigma$ values can be understood from the results given here. 
The evolution equation (\ref{feq2}) for the dimensionless Hubble parameter $h(z)$ becomes
\begin{align}\label{diffmain}
\frac{\textup{d}h}{\textup{d}z}=\frac{3}{2}\ \frac{h^2+\sigma\left(2\alpha-1\right)h^{2\alpha}}{\left[h+\sigma\alpha\left(2\alpha-1\right)h^{2\alpha-1}\right]\left(1+z\right)}\ .
\end{align}
Upon solving (\ref{diffmain}) for the dimensionless Hubble parameter, we turn our attention to studying the weak and strong focusing conditions for this cosmological model. In terms of $h(z)$, the weak focusing condition (\ref{weakonepara}) now reads
\begin{align}
 \label{wfc1}
\frac{8\pi\rho_{eff}}{3H_0^2}&=\left(\frac{1}{1+\sigma\alpha h^{2\alpha-2}}\right)\bigg[\Omega_{m_0}(1+z)^3+\sigma\left(1-\alpha\right)h^{2\alpha}\nonumber\\ &-\frac{\beta^2}{\left(1+z\right)^2}\left(\frac{4}{3}\sigma\alpha\left(\alpha-1\right)h^{2\alpha-1}h'\left(1+z\right)+h^2+\sigma h^{2\alpha}\right)\bigg]+\frac{\beta^2h^2}{\left(1+z\right)^2}\geq0\ .
\end{align}
where $h'$ denotes the derivative of the dimensionless Hubble parameter with respect to the redshift. In addition, for this cosmological model, the strong focusing condition  (\ref{strongonepara}) yields
\begin{align}
\frac{4\pi\left(\rho_{eff}+3P_{eff}\right)}{3H_0^2}&=\left(\frac{1}{1+\sigma\alpha h^{2\alpha-2}}\right)\bigg[\frac{1}{2}\Omega_{m_0}\left(1+z\right)^3-2\sigma\alpha\left(\alpha-1\right)h^{2\alpha-1}h'\left(1+z\right)\nonumber\\ &-\sigma\left(1-\alpha\right)h^{2\alpha}-\frac{\beta^2}{\left(1+z\right)^2}\left(\frac{4}{3}\sigma\alpha\left(\alpha-1\right)h^{2\alpha-1}h'\left(1+z\right)+h^2+\sigma h^{2\alpha}\right)\bigg]\nonumber\\ &+\frac{\beta^2h^2}{\left(1+z\right)^2}\geq0\ .
 \label{sfc1}
\end{align}
As already mentioned, $\Omega_{m_0}$ is dependent upon $\alpha$ and $\sigma$ through (\ref{feq1}). More specifically for this model, by evaluating (\ref{feq1}) today, i.e., at $z=0$, one gets
\begin{align}
\Omega_{m0}=h_0^2+\sigma\left(2\alpha-1\right)h_0^{2\alpha}\ ,
\end{align}
where $h_0:=h(z=0)$. We note that, since we have taken the Hubble parameter $H_0$ to be the value associated with the $\Lambda$CDM solution, we have $h_\Lambda(z=0)=1$. The value for $h_0$ is determined by the parameters $\alpha$ and $\sigma$. It is noted that one can make use of various data sets, such as cosmic microwave background data, in order to constrain the $f(T)$ model and therefore constrain $h_0$ and $\Omega_{m0}$. For the latest constraints placed on $f(T)$ cosmological models using various data sets, the reader is directed to \cite{Briffa:2020qli,Benetti:2020hxp}.

\begin{figure*}[!htb]
\minipage{0.32\textwidth}
  \includegraphics[width=\linewidth]{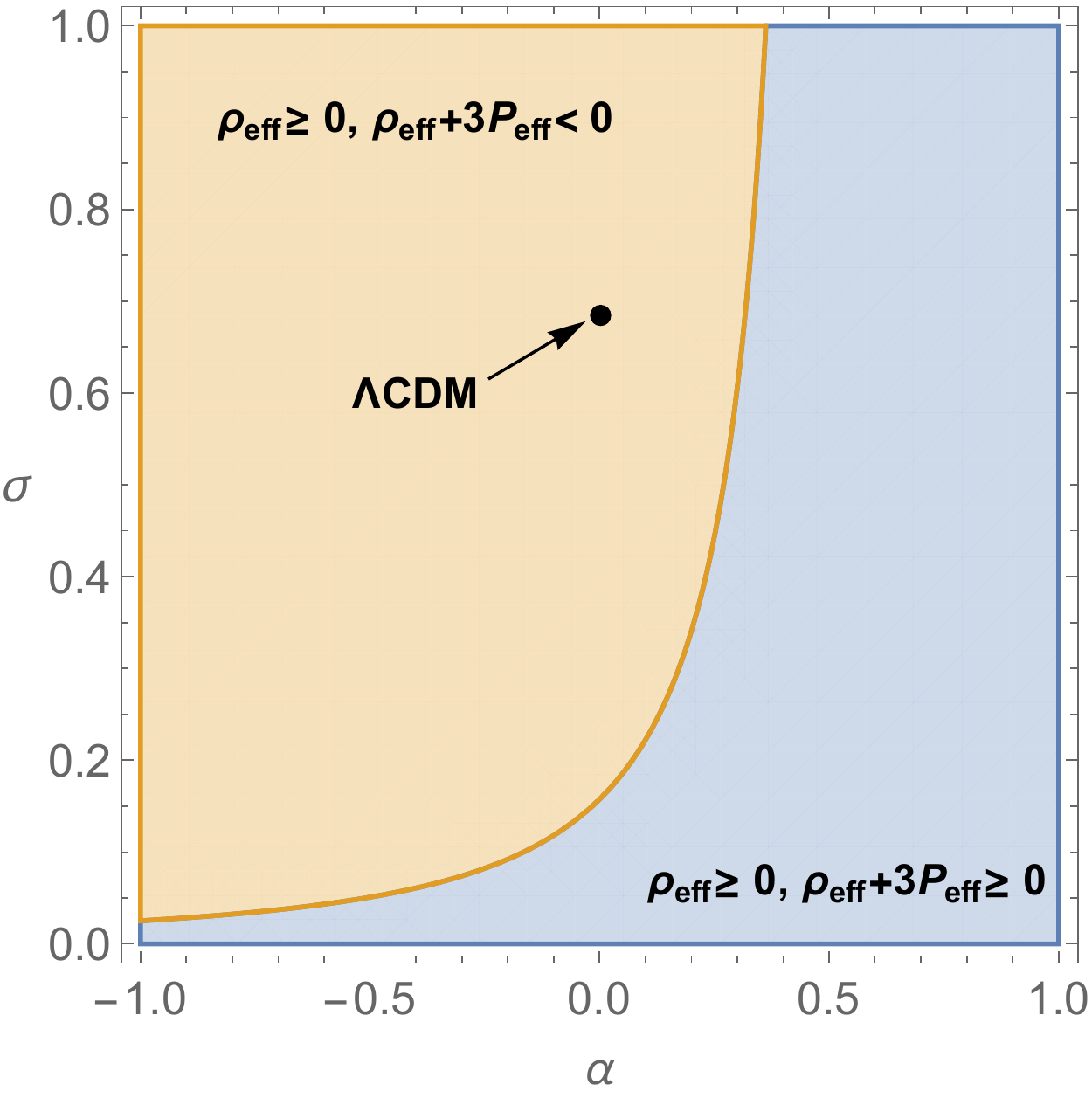}
\endminipage\hfill
\minipage{0.32\textwidth}
  \includegraphics[width=\linewidth]{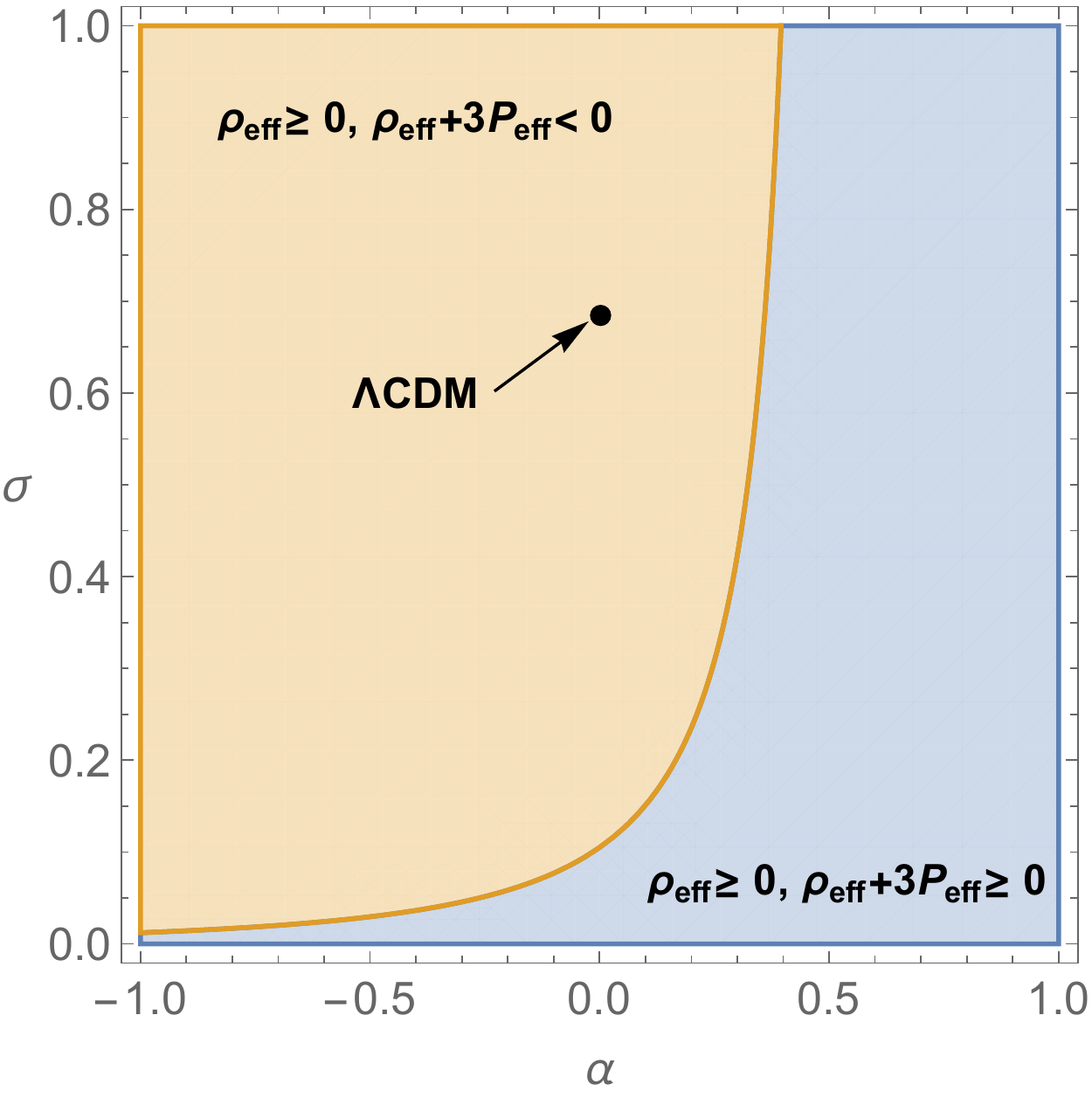}
\endminipage\hfill
\minipage{0.32\textwidth}%
  \includegraphics[width=\linewidth]{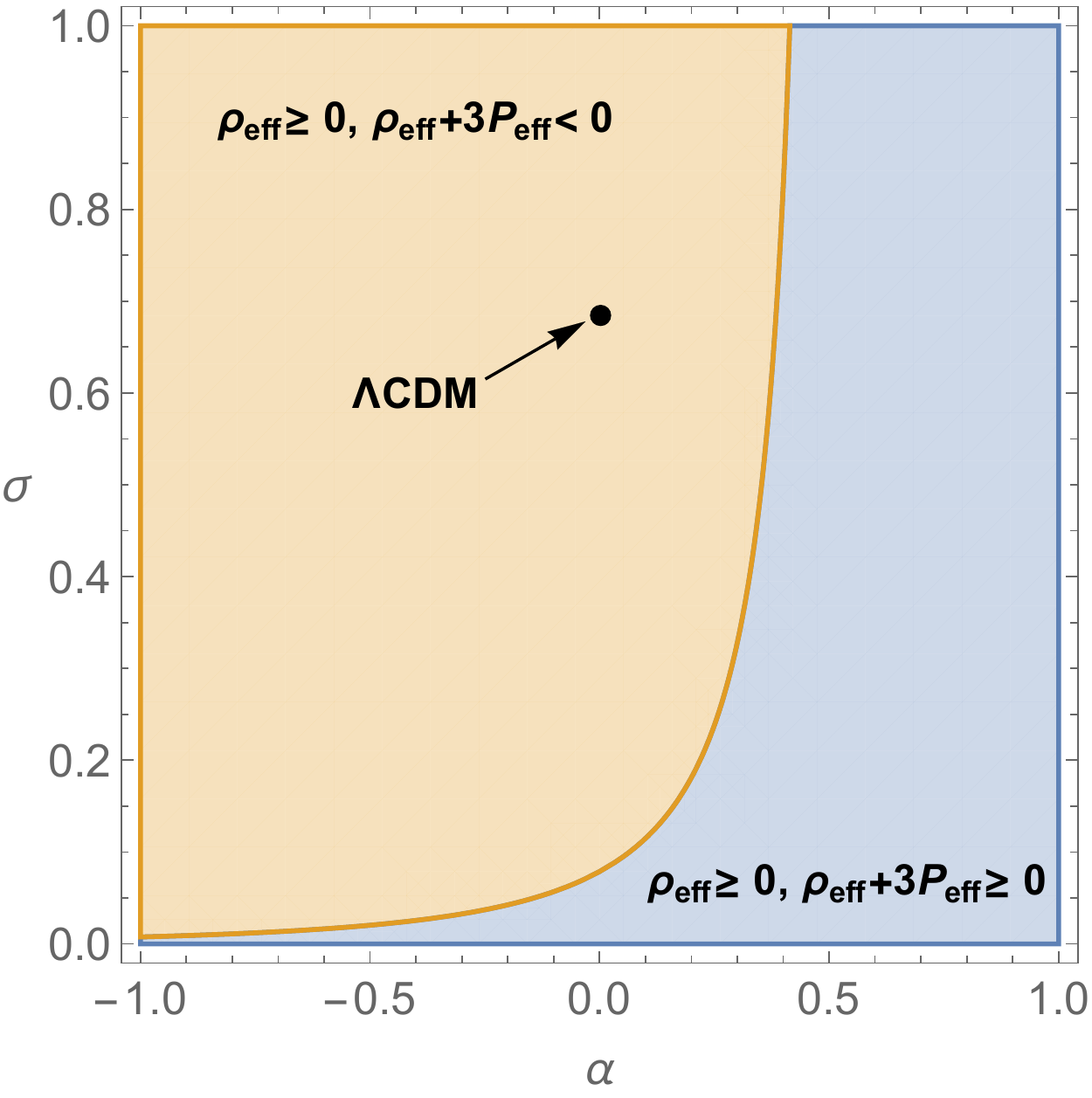}
\endminipage
\caption{Region plots showing the satisfaction or violation of the weak and strong focusing conditions for the polynomial cosmological model $f=6H_0^2\left(\bar{T}+\sigma\bar{T}^\alpha\right)$. The blue regions indicate where both the weak and the strong focusing conditions are satisfied whereas the orange regions indicate where only the weak focusing condition is satisfied. The left, middle and right panels show the region plots for the cases where $C=0$, $C=1$ and $C=10^4$ respectively. We note that the weak focusing condition is satisfied for all values of $C$ considered. In addition, we note that when the value of $C$ is increased, the parameter space area violating the strong focusing condition is bigger. In these plots, we have indicated where the $\Lambda$CDM solution appears by plotting it as a point.}
\label{fig: rpol}
\end{figure*}

\begin{figure*}[!htb]
\minipage{0.32\textwidth}
  \includegraphics[width=\linewidth]{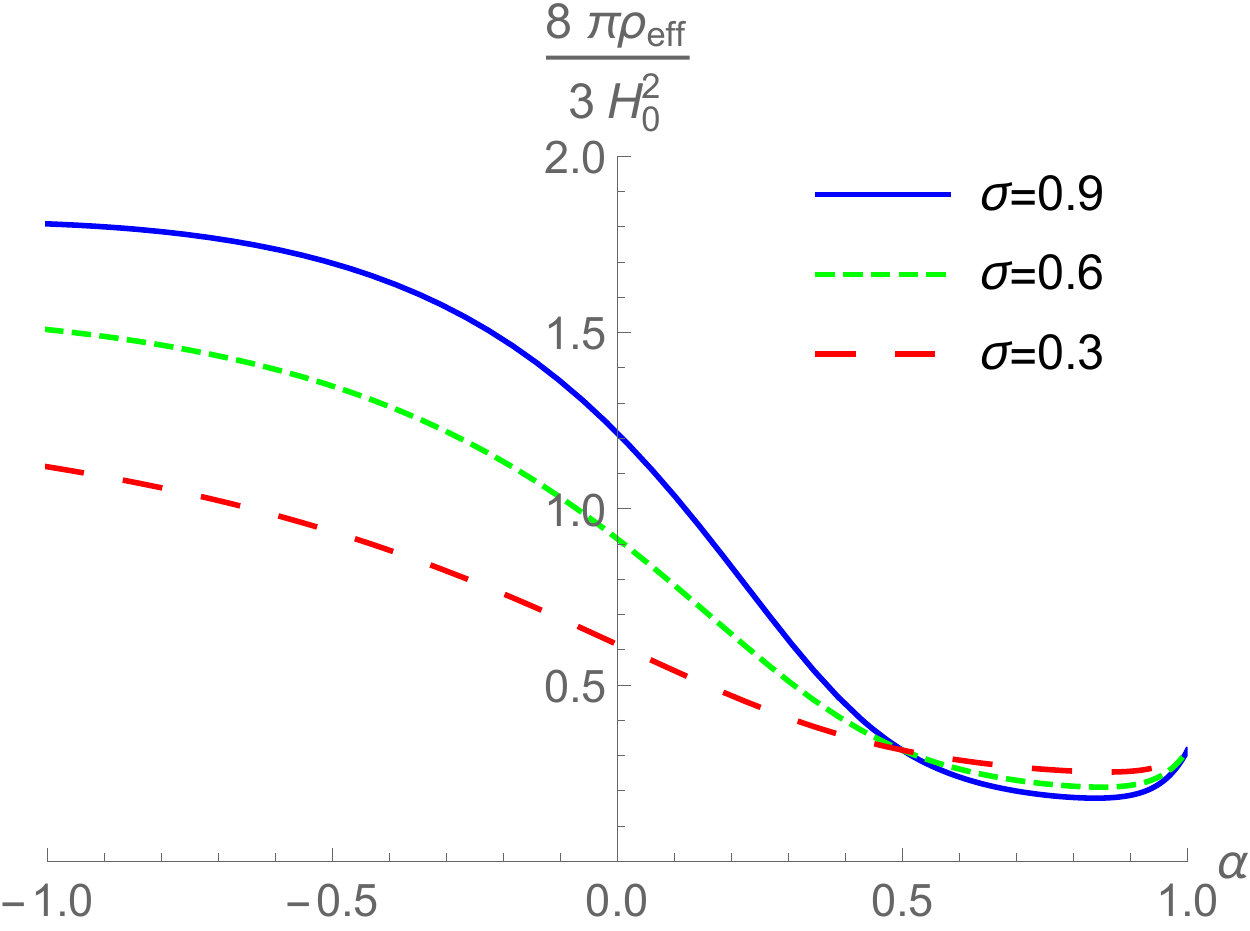}
\endminipage\hfill
\minipage{0.32\textwidth}
  \includegraphics[width=\linewidth]{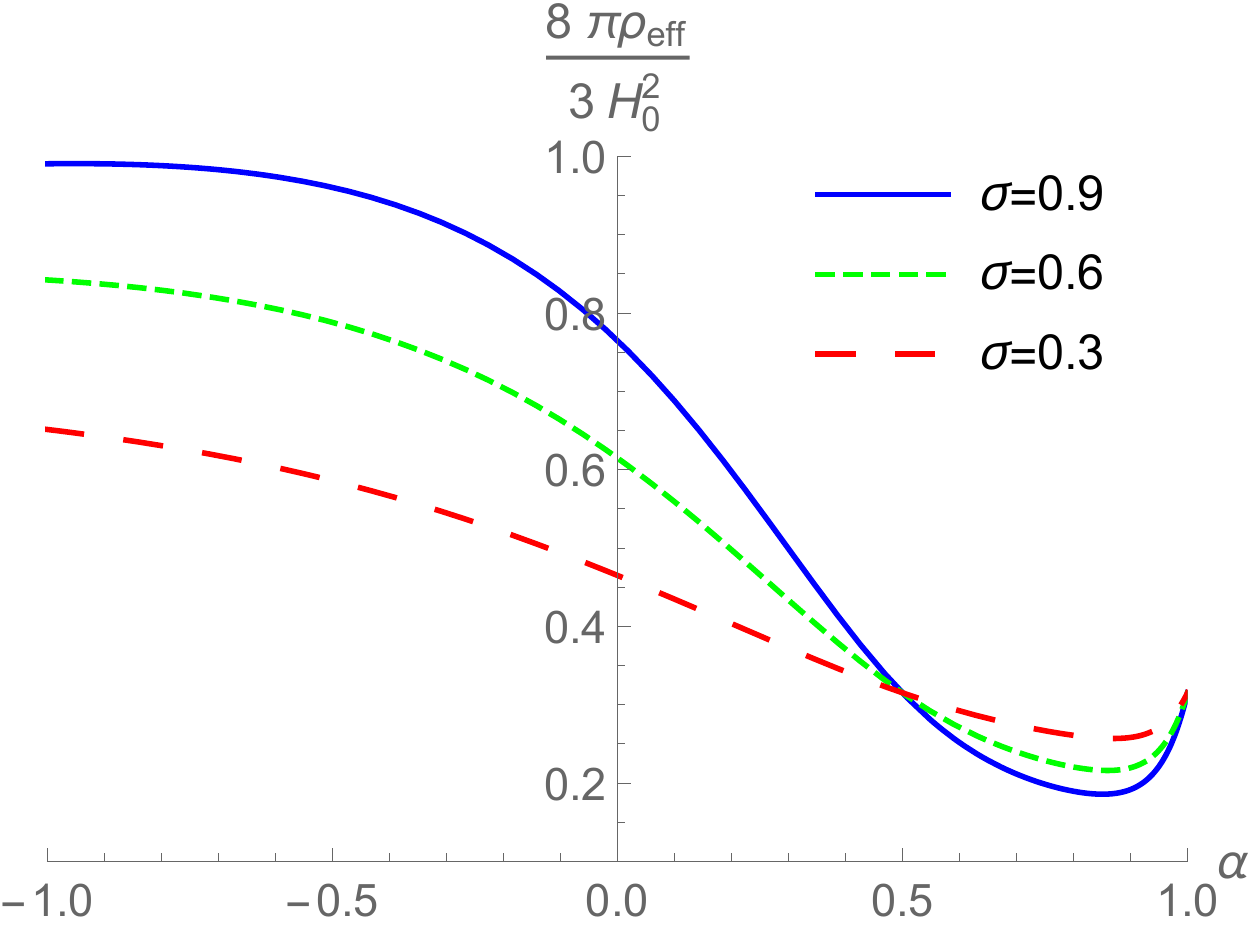}
\endminipage\hfill
\minipage{0.32\textwidth}%
  \includegraphics[width=\linewidth]{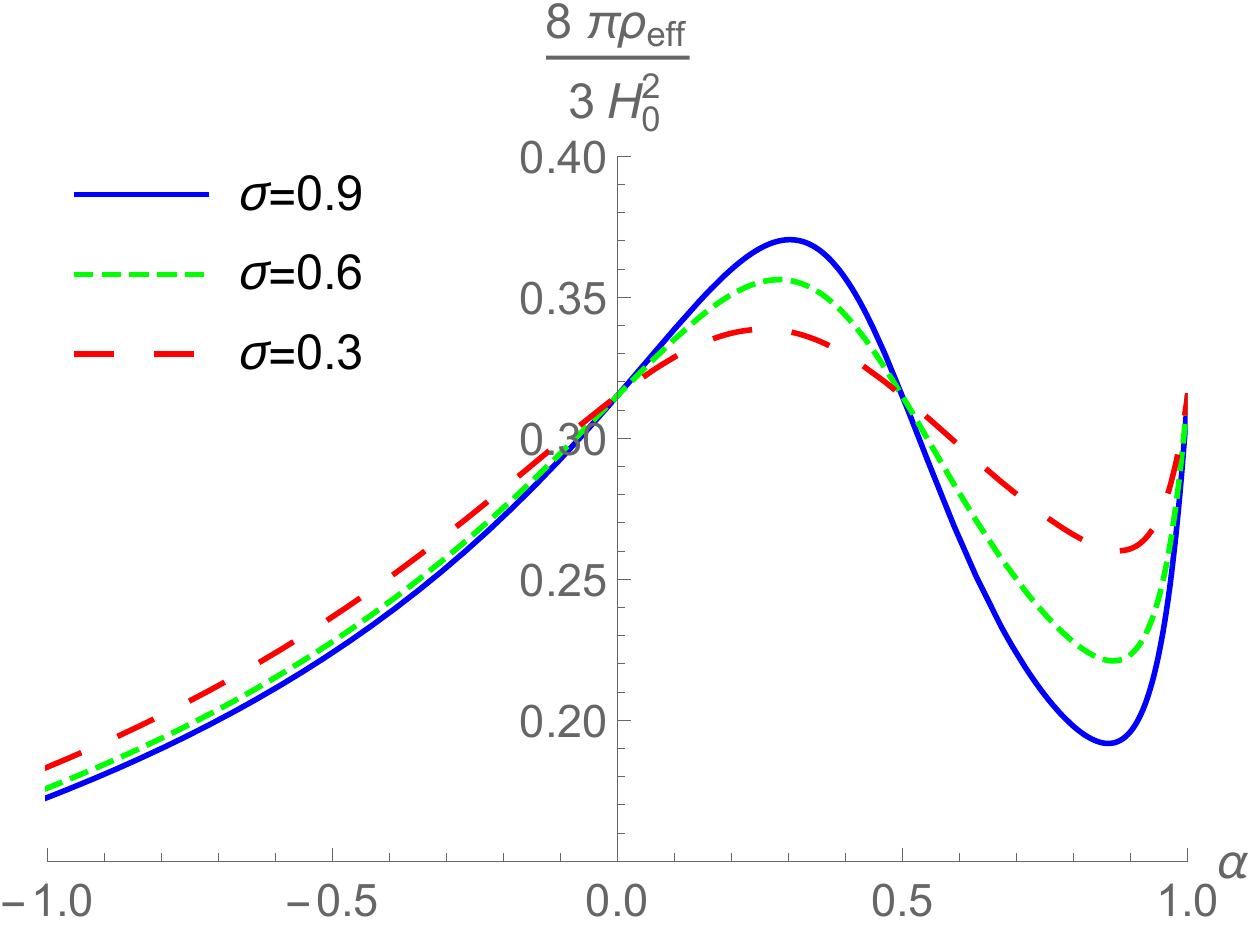}
\endminipage
\caption{$\sigma$-constant profile plots of equation (\ref{wfc1}) for the polynomial cosmological model $f=6H_0^2\left(\bar{T}+\sigma\bar{T}^\alpha\right)$. The solid blue curves show the $\sigma=0.9$ profiles, the dotted green curves show the $\sigma=0.6$ profiles and the dashed red curves show the $\sigma=0.3$ profiles. The left, middle and right panels correspond to $C=0$, $C=1$ and $C=10^4$ respectively.}
\label{fig: wpol}
\end{figure*}

\begin{figure*}[!htb]
\minipage{0.32\textwidth}
  \includegraphics[width=\linewidth]{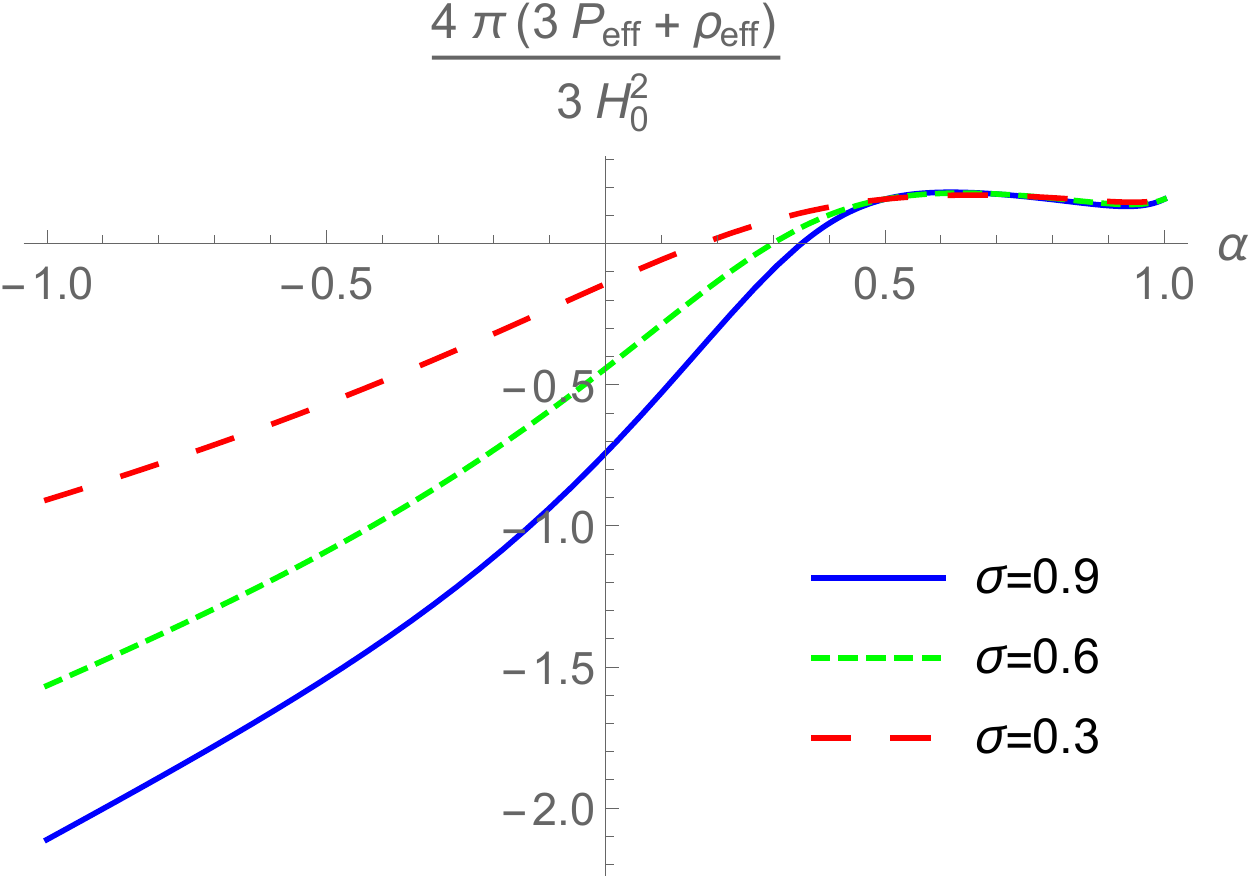}
\endminipage\hfill
\minipage{0.32\textwidth}
  \includegraphics[width=\linewidth]{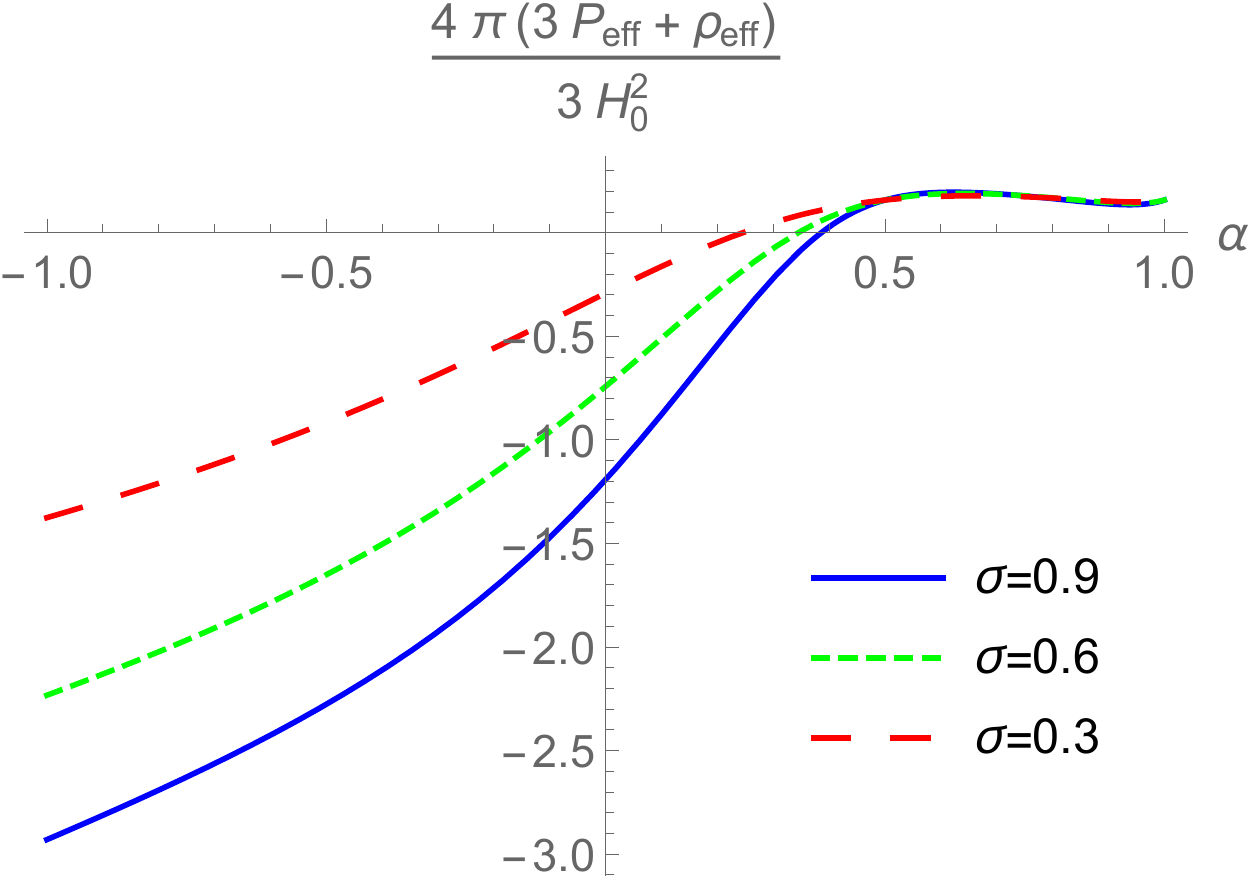}
\endminipage\hfill
\minipage{0.32\textwidth}%
  \includegraphics[width=\linewidth]{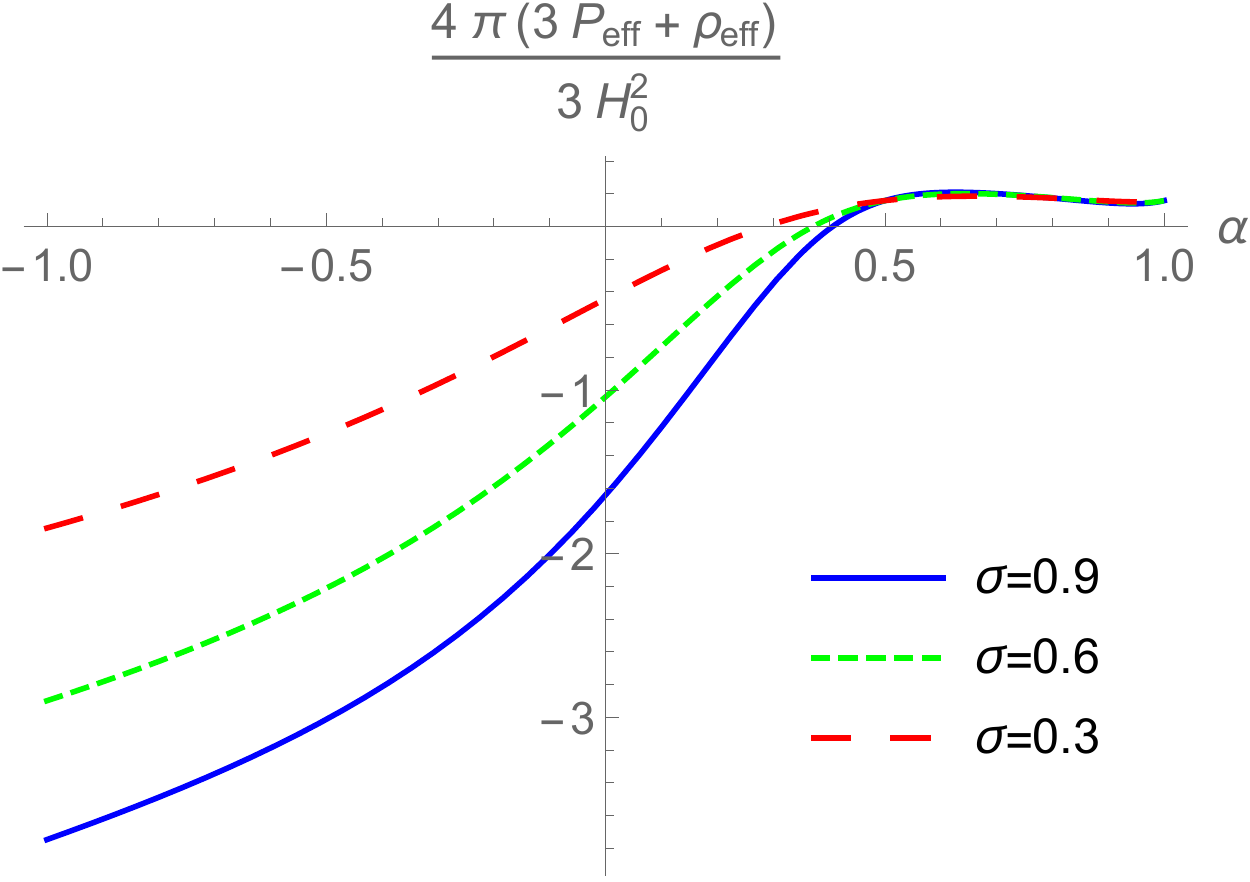}
\endminipage
\caption{$\sigma$-constant profiles of equation (\ref{sfc1}) for the polynomial cosmological model $f=6H_0^2\left(\bar{T}+\sigma\bar{T}^\alpha\right)$. The solid blue curves show the $\sigma=0.9$ profiles, the dotted green curves show the $\sigma=0.6$ profiles and the dashed red curves show the $\sigma=0.3$ profiles. The left, middle and right panels correspond to $C=0$, $C=1$ and $C=10^4$ respectively.}
\label{fig: spol}
\end{figure*}

Let us now consider the $\{\alpha,\,\sigma\}$ regions for which the weak and strong focusing conditions, as provided in (\ref{wfc1}) and (\ref{sfc1}) respectively, are satisfied or violated\footnote{In this note, all plots are produced using the present-day value, i.e., $z=0$.}. 
Figure \ref{fig: rpol} contains region plots showing the satisfaction or violation of the weak and strong focusing conditions in the aforementioned $\{\alpha,\,\sigma\}$ parameter space. The blue regions indicate where both the weak and strong focusing conditions are satisfied, whereas the orange regions show where only the weak focusing condition is satisfied. The first panel shows the region plot for the case of a fundamental congruence, i.e., when $C=0$. The second panel shows the region plot for the case where $C=1$ and the third panel shows the region plot associated with the case where $C=10^4$. 
It is evident from Figure \ref{fig: rpol} that there exist $\{\alpha,\,\sigma\}$ combinations allowing for the strong focusing condition to be violated. 
In addition, Figure \ref{fig: rpol} illustrates how the satisfaction of the strong focusing condition changes as the value for $C$ changes, i.e., when considering different four-velocities of the form given in equation (\ref{onepara}). As the value for $C$ is increased,  the $\{\alpha,\,\sigma\}$ parameter space area violating the strong focusing condition gets bigger. 
For all cases considered, the weak focusing condition remains satisfied. 
In order to illustrate where in these contour plots the $\Lambda$CDM solution apears, we have plotted it as a single point.
Figure \ref{fig: wpol} shows the profile plots of equation (\ref{wfc1}) for constant $\sigma$ values . The first, second and third panels show such profile plots for the cases $C=0$, $C=1$ and $C=10^4$ respectively. Analogously, the constant $\sigma$ profile plots of equation (\ref{sfc1}) are shown in Figure \ref{fig: spol}. For the latter scenario, as the value of $C$ increases, the limiting value of $\alpha$ at which the strong focusing condition is violated gets higher. Here, we do not give the constant $\alpha$ profiles since these are not as interesting due to the fact that $\sigma$ plays a role that is similar to that of a cosmological constant.

\subsection{Model 2: $f=6H_0^2\left(\bar{T}+\sigma\,{\rm e}^{\alpha\bar{T}}\right)$}
\label{Sec:7:A}

The second cosmological model considered is of the exponential form given above \cite{bamba}. Herein, we shall restrict our consideration to $\sigma$ and $\alpha$ values such that $\left(\sigma,\alpha\right)\in\left[0,{\rm e}\right]\times[-1,0]$. We note that the viability conditions are satisfied in this region. For the exponential cosmological model given above, equation (\ref{feq2}) for the case of vanishing pressure reads
\begin{align}\label{diffexp}
\frac{\textup{d}h}{\textup{d}z}=\frac{3}{2}\frac{h^2+\sigma {\rm{e}}^{\alpha h^2}\left(2\alpha h^2-1\right)}{\left(h+\sigma\alpha h\ {\rm{e}}^{\alpha h^2}+2\sigma\alpha^2h^3{\rm{e}}^{\alpha h^2}\right)\left(1+z\right)}\ .
\end{align}
Now, from equation (\ref{weakonepara}), the weak focusing condition for the exponential cosmological model reads
\begin{align}
\frac{8\pi\rho_{eff}}{3H_0^2}&=\frac{1}{\left(1+\sigma\alpha\ {\rm{e}}^{\alpha h^2}\right)}\bigg[\Omega_{m_0}\left(1+z\right)^3+\sigma {\rm{e}}^{\alpha h^2}\left(1-\alpha h^2\right)\nonumber \\
&-\frac{\beta^2}{\left(1+z\right)^2}\left(\frac{4}{3}h^3\left(1+z\right)h'\sigma\alpha^2 {\rm{e}}^{\alpha h^2}+h^2+\sigma {\rm{e}}^{\alpha h^2}\right)\bigg]+\frac{\beta^2h^2}{\left(1+z\right)^2}\geq0\ ,
\label{wfc2} 
\end{align}
where we obtain the values for $h(z)$ by solving the differential equation (\ref{diffexp}). In addition, from equation (\ref{strongonepara}), we find that the strong focusing condition for the exponential cosmological model reads
\begin{align}
\frac{4\pi\left(\rho_{eff}+3P_{eff}\right)}{3H_0^2}&=\frac{1}{\left(1+\sigma\alpha\ {\rm{e}}^{\alpha h^2}\right)}\bigg[\frac{1}{2}\Omega_{m_0}\left(1+z\right)^3-\sigma {\rm{e}}^{\alpha h^2}(1
-\alpha h^2)-2\left(1+z\right)h^3h'\sigma\alpha^2{\rm{e}}^{\alpha h^2}\nonumber\\ &-\frac{\beta^2}{\left(1+z\right)^2}\bigg(\frac{4}{3}h^3\left(1+z\right)h'\sigma\alpha^2 {\rm{e}}^{\alpha h^2}
+h^2+\sigma {\rm{e}}^{\alpha h^2}\bigg)\bigg]+\frac{\beta^2h^2}{\left(1+z\right)^2}\geq0\ .
\label{sfc2}
\end{align}
As done for the case of the polynomial model, the energy density parameter $\Omega_{m_0}$  depends on the values of $\sigma$ and $\alpha$. More specifically, by evaluating (\ref{feq1}) at $h(z=0)=h_0$, $\Omega_{m_0}$ yields
\begin{align}
\Omega_{m_0}=h_0^2+\sigma {\rm{e}}^{\alpha h_0^2}\left(2\alpha h^2_0-1\right)\ .
\end{align}

\begin{figure*}[!htb]
\minipage{0.32\textwidth}
  \includegraphics[width=\linewidth]{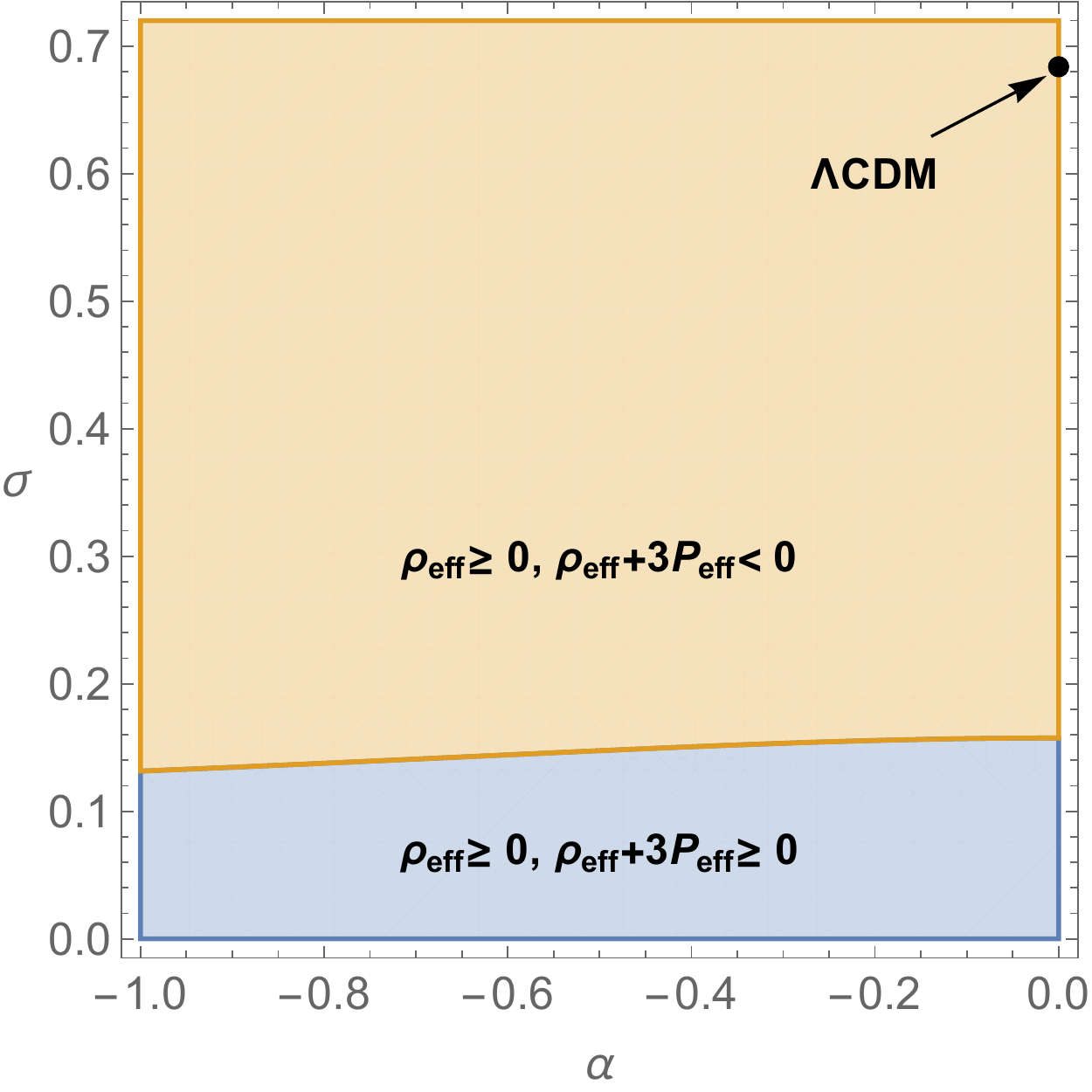}
\endminipage\hfill
\minipage{0.32\textwidth}
  \includegraphics[width=\linewidth]{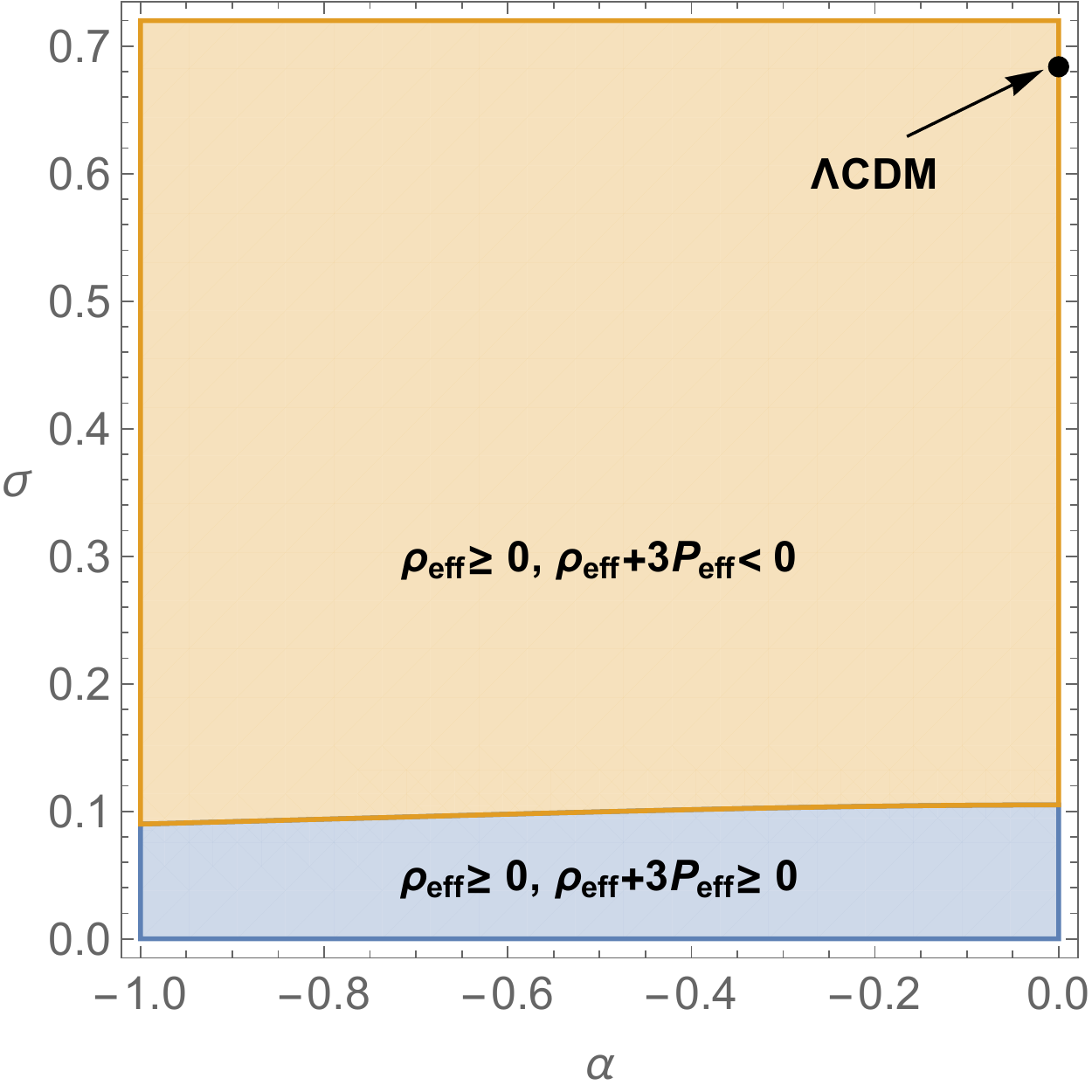}
\endminipage\hfill
\minipage{0.32\textwidth}%
  \includegraphics[width=\linewidth]{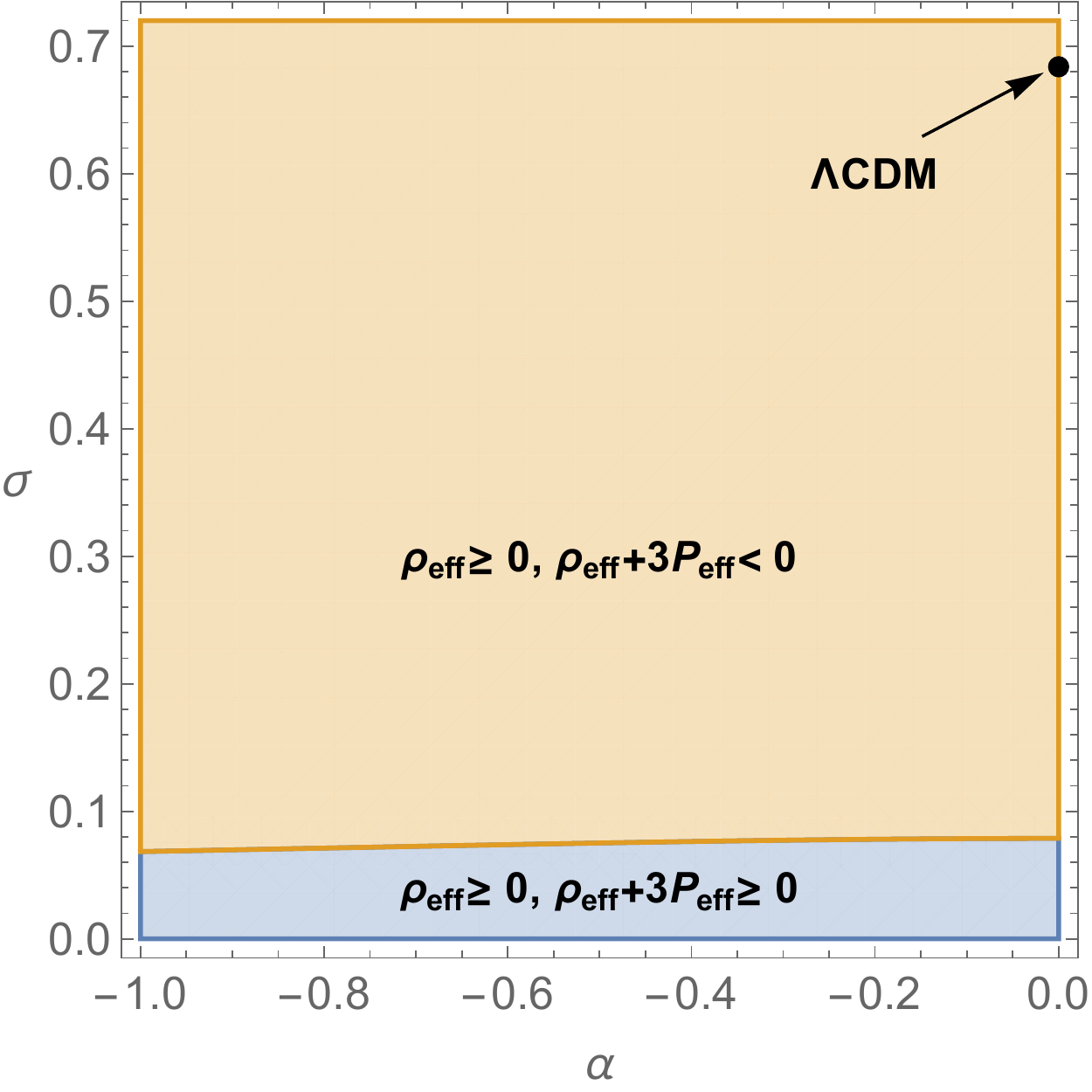}
\endminipage
\caption{Region plots showing the satisfaction or violation of the weak and strong focusing conditions for the exponential cosmological model $f=6H_0^2\left(\bar{T}+\sigma{\rm e}^{\alpha\bar{T}}\right)$. The blue regions indicate where both the weak and the strong focusing conditions are satisfied whereas the orange regions indicate where only the weak focusing condition is satisfied. The left, middle and right panels show the region plots for the cases of $C=0$, $C=1$ and where $C=10^4$ respectively. The location of the $\Lambda$CDM solution is plotted as a point. We note that the weak focusing condition is satisfied for all values of $C$ considered. In addition, 
 we note that when the value of $C$ is increased, the parameter space area violating the strong focusing condition is bigger.}
\label{fig: rexp}
\end{figure*}

\begin{figure*}[!htb]
\minipage{0.32\textwidth}
  \includegraphics[width=\linewidth]{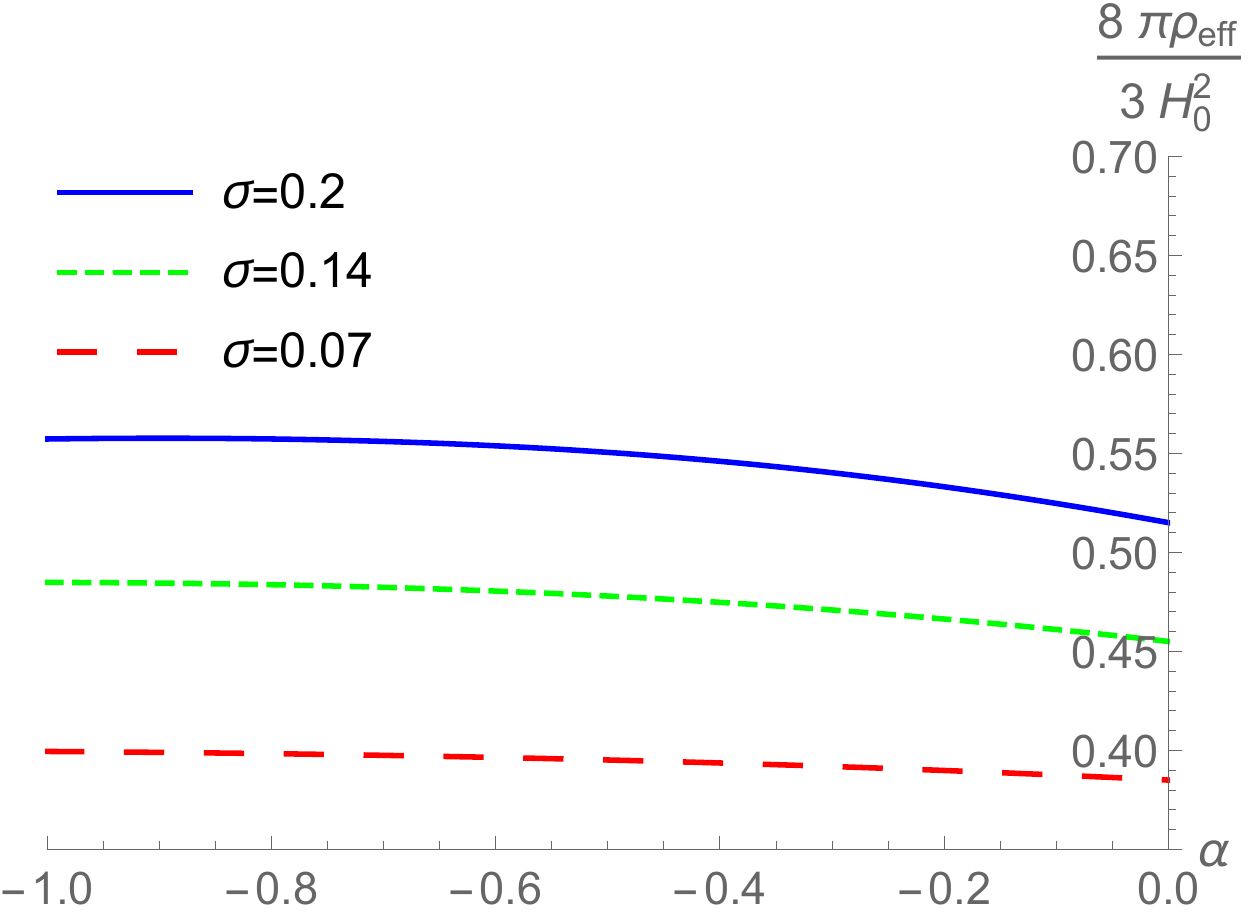}
\endminipage\hfill
\minipage{0.32\textwidth}
  \includegraphics[width=\linewidth]{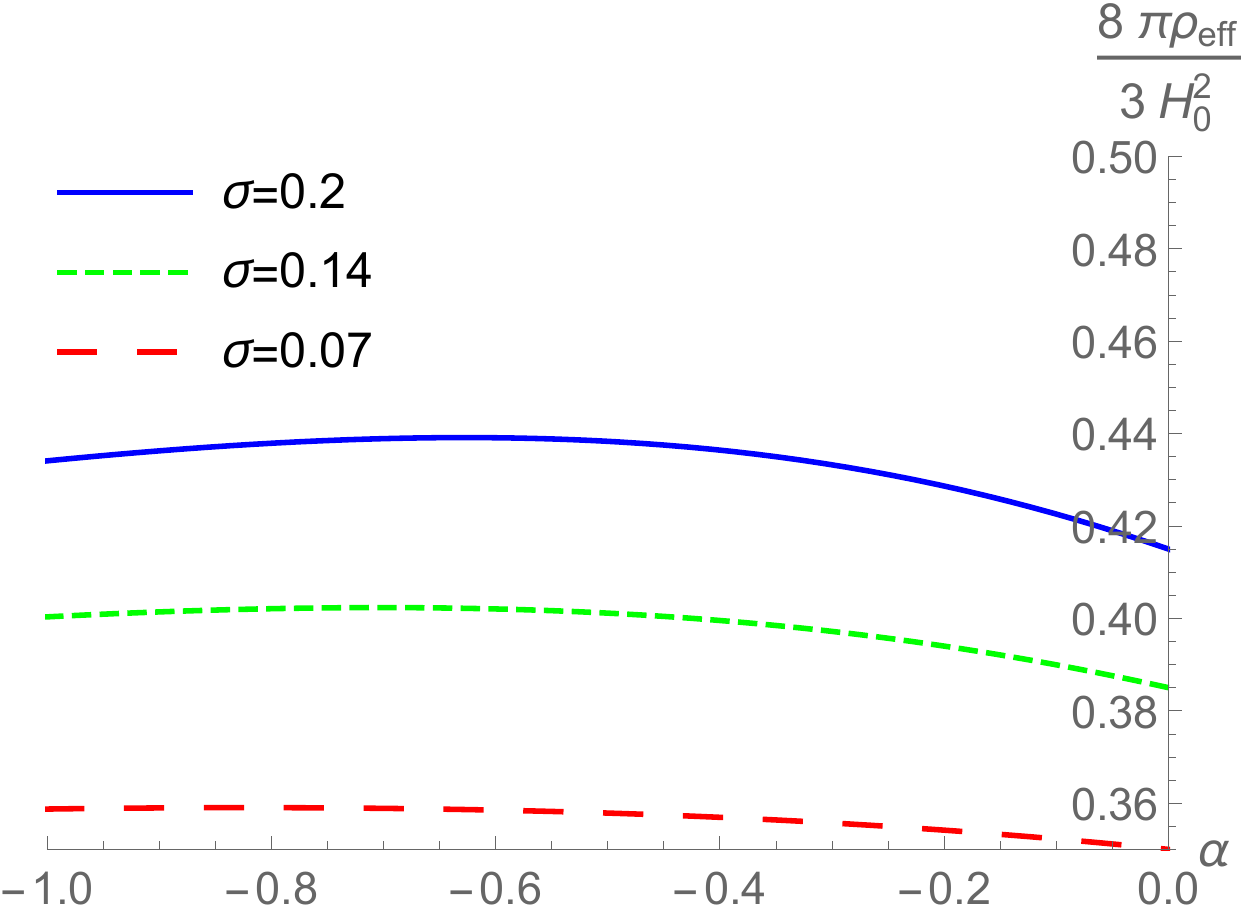}
\endminipage\hfill
\minipage{0.32\textwidth}%
  \includegraphics[width=\linewidth]{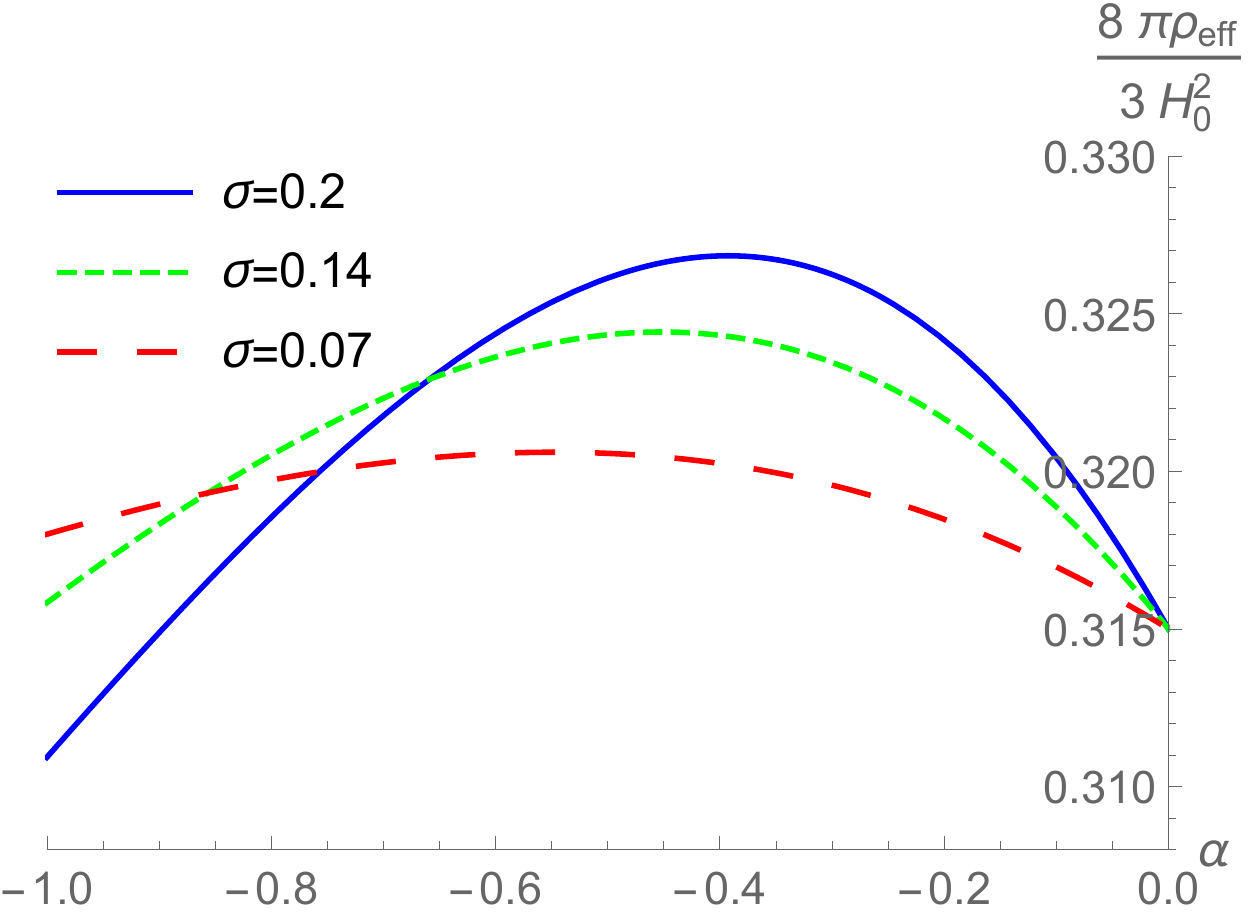}
\endminipage
\caption{$\sigma$-constant profiles of equation (\ref{wfc2}) for the exponential cosmological model $f=6H_0^2\left(\bar{T}+\sigma {\rm e}^{\alpha\bar{T}}\right)$. The solid blue curves show the $\sigma=0.2$ profiles, the dotted green curves show the $\sigma=0.14$ profiles and the dashed red curves show the $\sigma=0.07$ profiles. The left, middle and right panels correspond to $C=0$, $C=1$ and $C=10^4$ respectively.}
\label{fig: wexp}
\end{figure*}

\begin{figure*}[!htb]
\minipage{0.32\textwidth}
  \includegraphics[width=\linewidth]{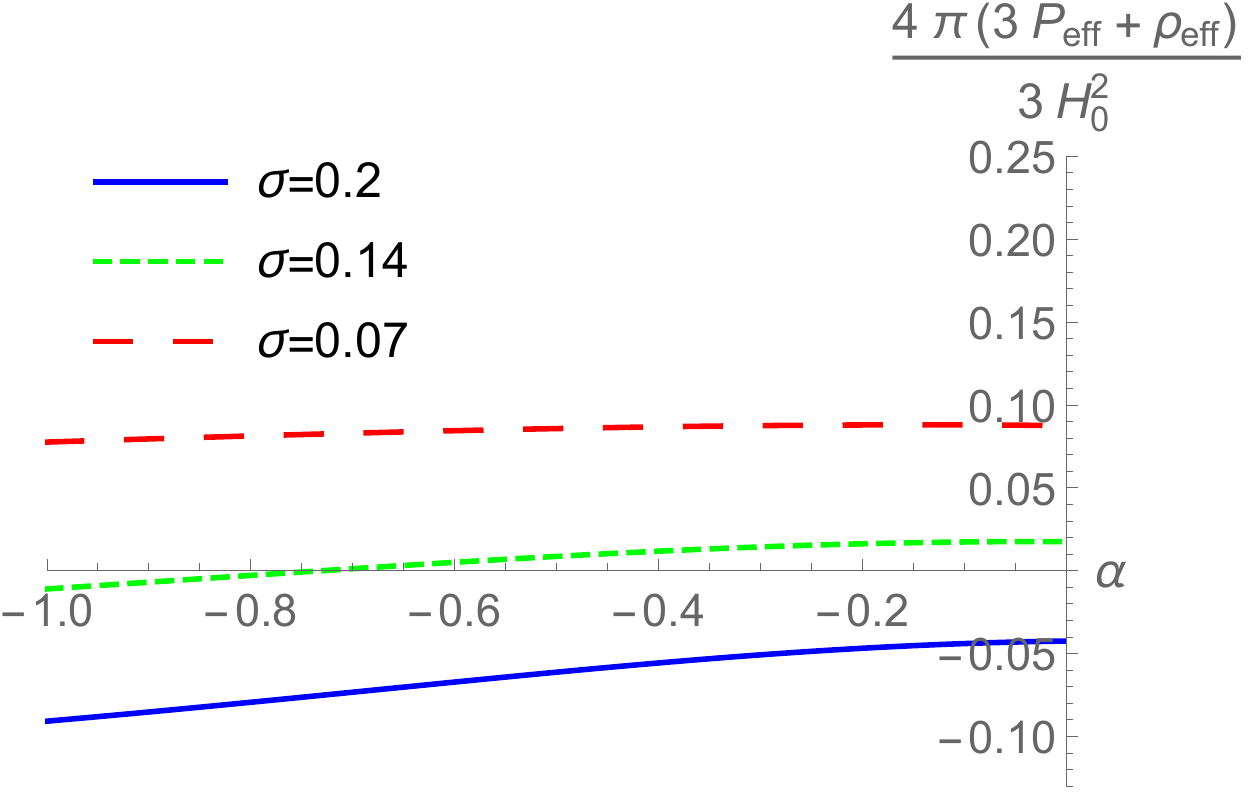}
\endminipage\hfill
\minipage{0.32\textwidth}
  \includegraphics[width=\linewidth]{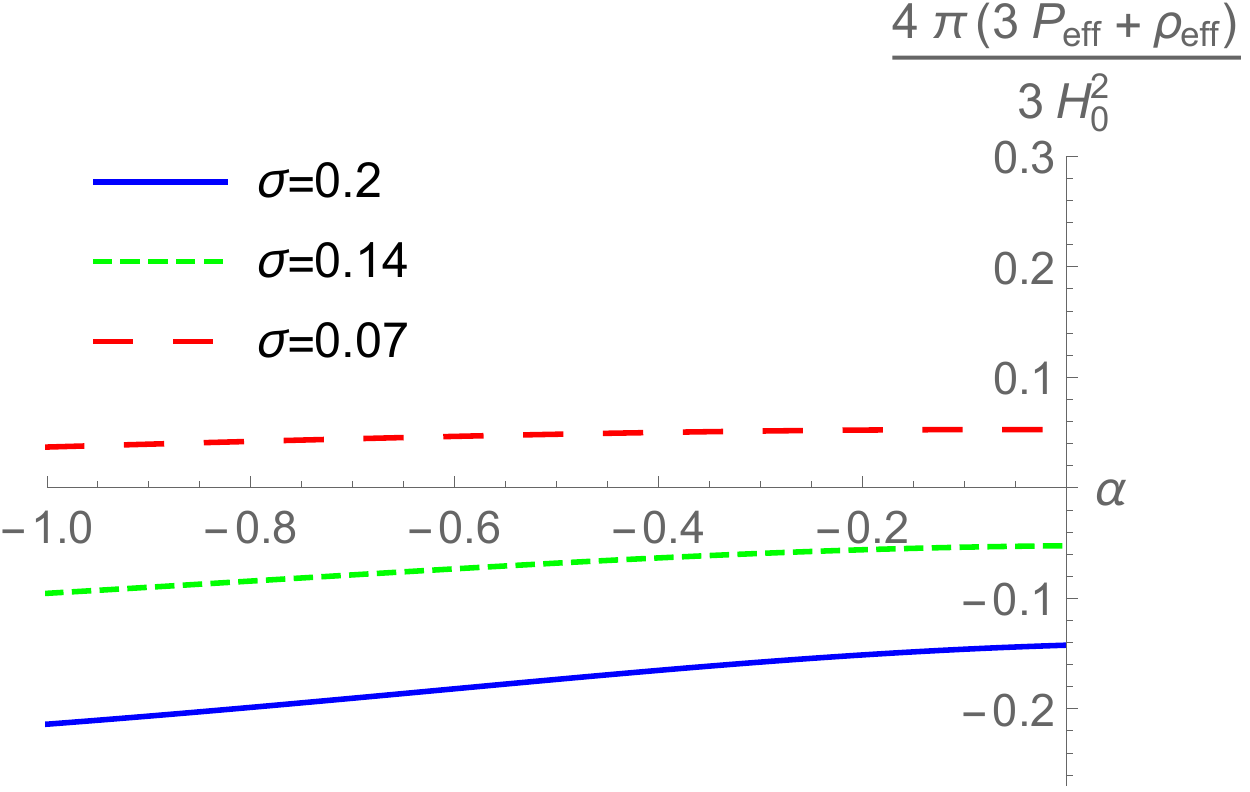}
\endminipage\hfill
\minipage{0.32\textwidth}%
  \includegraphics[width=\linewidth]{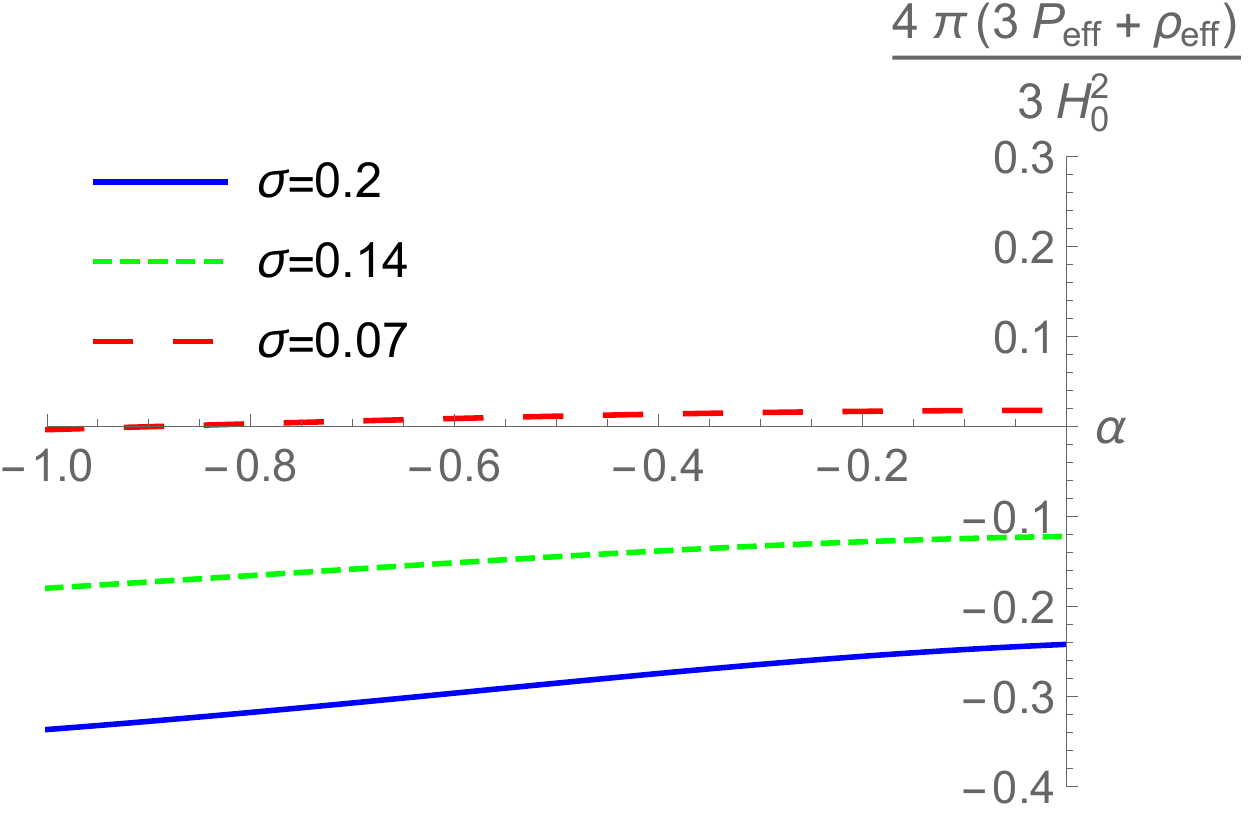}
\endminipage
\caption{$\sigma$-constant profiles of equation (\ref{sfc2}) for the exponential cosmological model $f=6H_0^2\left(\bar{T}+\sigma {\rm e}^{\alpha\bar{T}}\right)$. The solid blue curves show the $\sigma=0.2$ profiles, the dotted green curves show the $\sigma=0.14$ profiles and the dashed red curves show the $\sigma=0.07$ profiles. The left, middle and right panels correspond to $C=0$, $C=1$ and $C=10^4$ respectively.}
\label{fig: sexp}
\end{figure*}

Analogously to the first model above, the satisfaction or violation of the weak and strong focusing conditions for this model in the $\{\alpha,\, \sigma\}$ parameter space can be studied using
(\ref{wfc2}) and (\ref{sfc2}) respectively.
Thus, Figure \ref{fig: rexp} shows the region plots of the satisfaction or violation of the weak and strong focusing conditions. The blue regions indicate where both the weak and strong focusing conditions are satisfied. The orange regions indicate where only the strong focusing condition is violated. There, the first, second and third panels are produced by considering the cases where $C=0$, $C=1$ and $C=10^4$ respectively. As observed in the polynomial model above, when the value of $C$ is increased, the $\{\alpha,\, \sigma\}$ parameter space where the strong focusing condition is violated increases. Also, for this cosmological model, the weak focusing condition remains satisfied. Figures \ref{fig: wexp} and \ref{fig: sexp} show constant $\sigma$ profiles of equations (\ref{wfc2}) and (\ref{sfc2}) respectively.
A comment worthwhile to be made concerns the constant $\sigma$ profiles in Figure \ref{fig: sexp}. As the value of $C$ increases (from left to right  panels) and for a fixed $\sigma$, the left-hand side of equation (\ref{sfc2}) seems to take smaller values which could be eventually negative. Therefore  the strong focusing condition may be violated.


\subsection{Model 3: $f=6H_0^2\Big(\bar{T}+\sigma\bar{T}^\alpha\tanh\big(1/\bar{T}\big)\Big)$}
\label{Sec:7:A}

%(\ref{wfc2}) and (\ref{sfc2}) respectively.
% \label{wfc2}  %\label{sfc2}

\begin{figure*}[!htb]
\minipage{0.32\textwidth}
  \includegraphics[width=\linewidth]{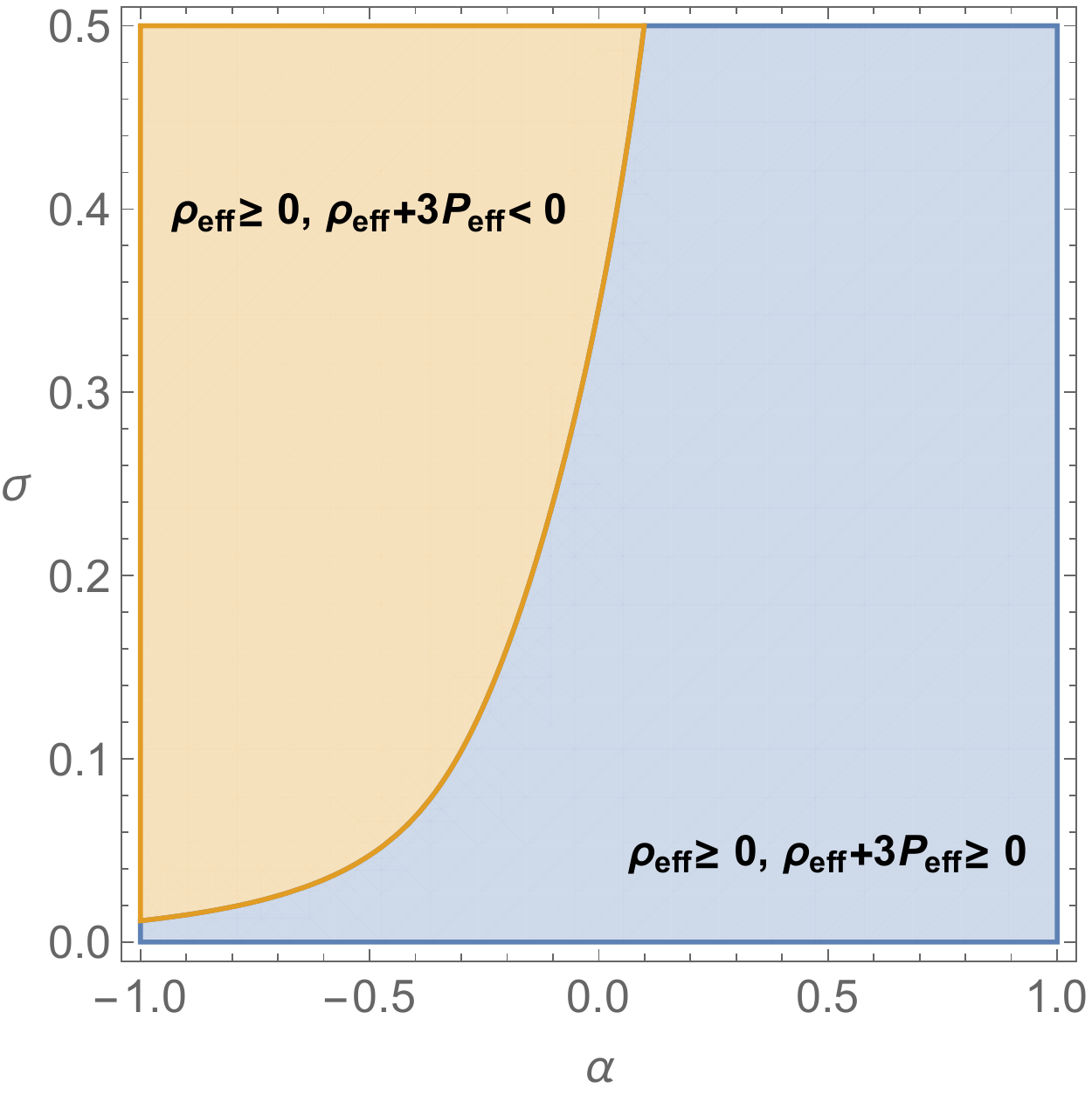}
\endminipage\hfill
\minipage{0.32\textwidth}
  \includegraphics[width=\linewidth]{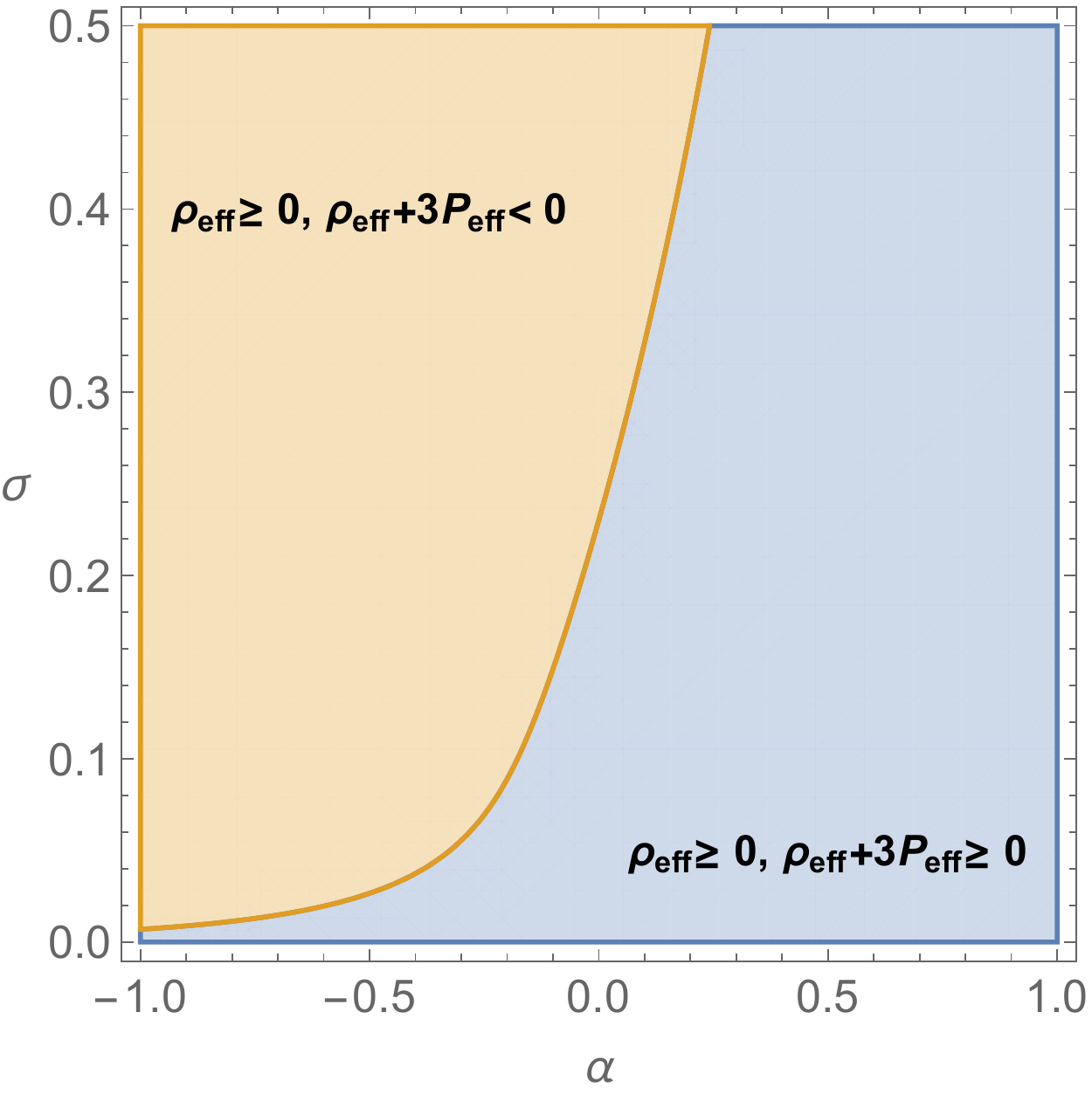}
\endminipage\hfill
\minipage{0.32\textwidth}%
  \includegraphics[width=\linewidth]{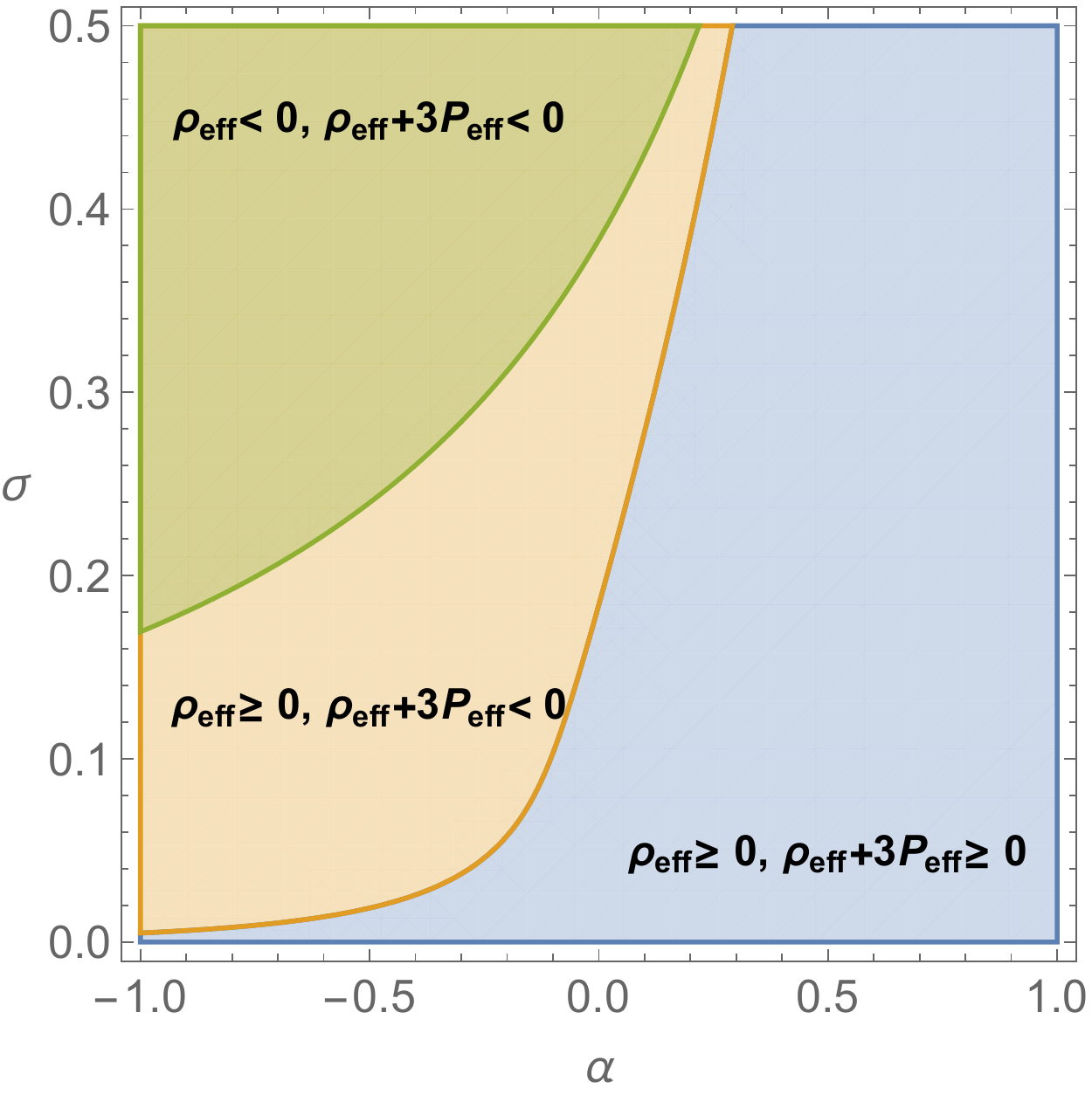}
\endminipage
\caption{Region plots showing the satisfaction or violation of the weak and strong focusing conditions for the hyperbolic tangent cosmological model $f=6H_0^2\left(\bar{T}+\sigma\bar{T}^\alpha\tanh\left(1/\bar{T}\right)\right)$. The blue regions indicate where both the weak and the strong focusing conditions are satisfied whereas the orange regions indicate where only the weak focusing condition is satisfied. The green region indicates where both the weak and the strong focusing conditions are violated. The left, middle and right panels show the region plots for the cases of $C=0$, $C=1$ and $C=10^4$ respectively. We note that more regions for which the strong focusing condition is violated arise when the value for $C$ is increased. In addition, we note that, while the weak focusing condition is satisfied for the case of a fundamental congruence, regions for which the weak focusing condition is violated arise by increasing the value of $C$.}
\label{fig: rtanh}
\end{figure*}

\begin{figure*}[!htb]
\minipage{0.32\textwidth}
  \includegraphics[width=\linewidth]{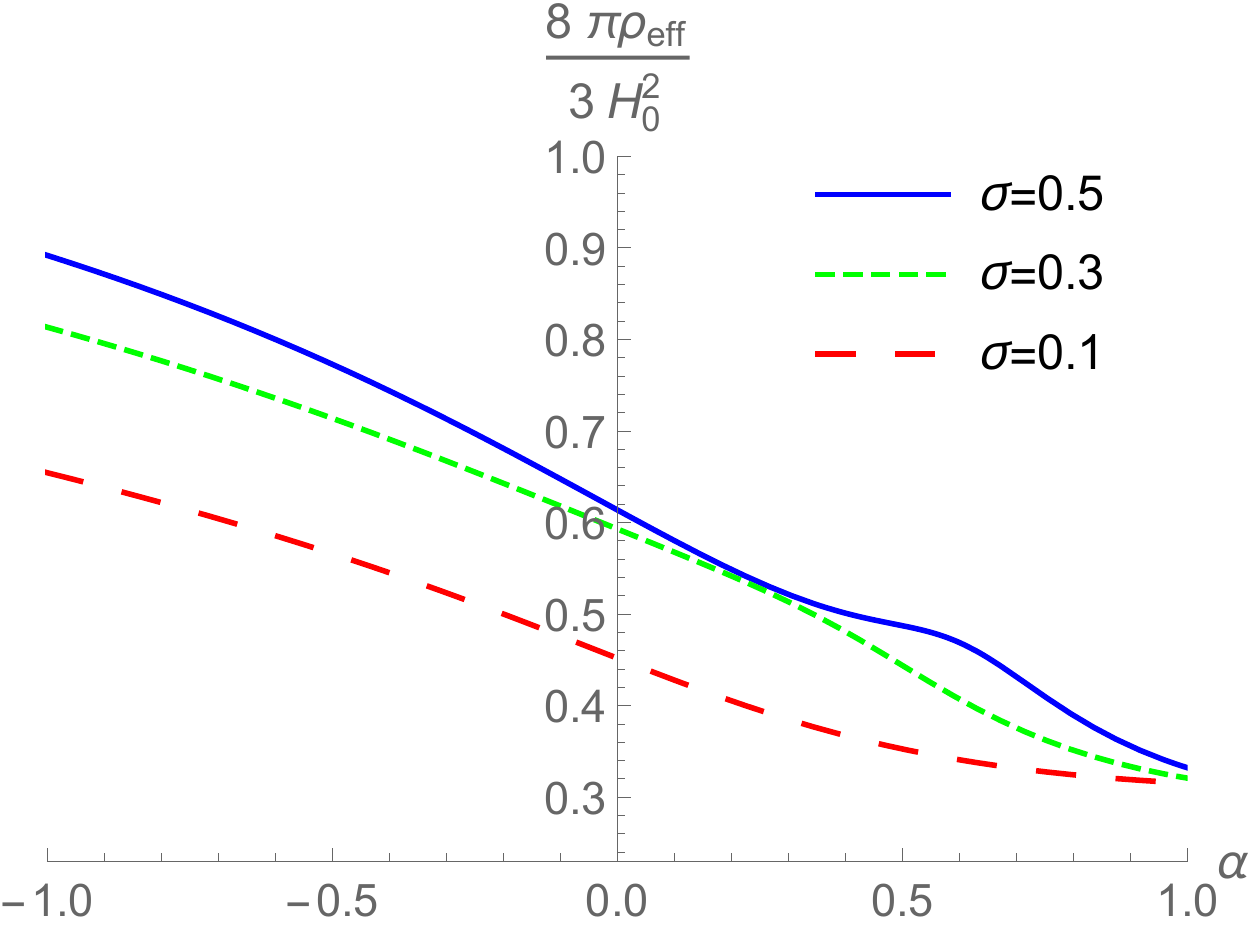}
\endminipage\hfill
\minipage{0.32\textwidth}
  \includegraphics[width=\linewidth]{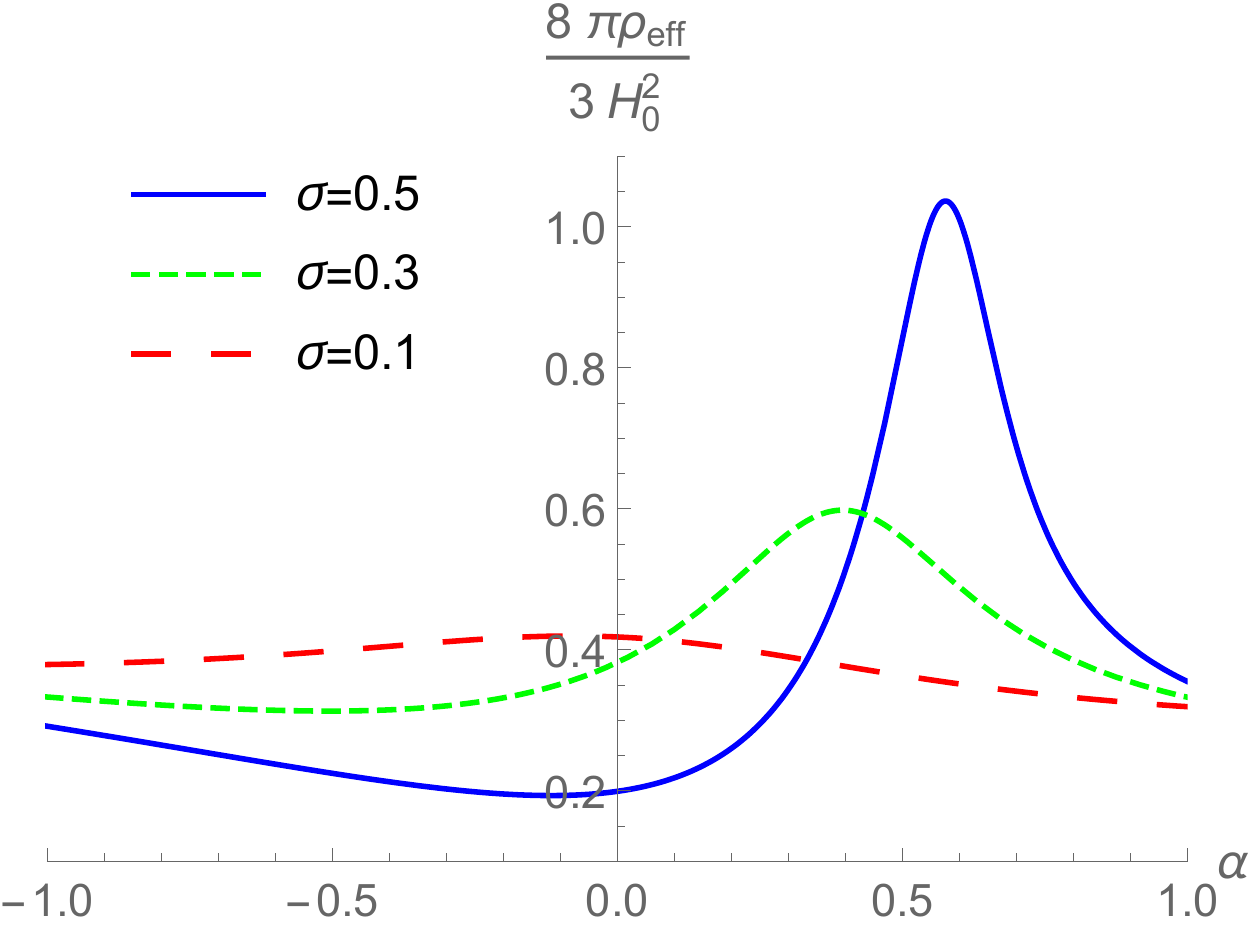}
\endminipage\hfill
\minipage{0.32\textwidth}%
  \includegraphics[width=\linewidth]{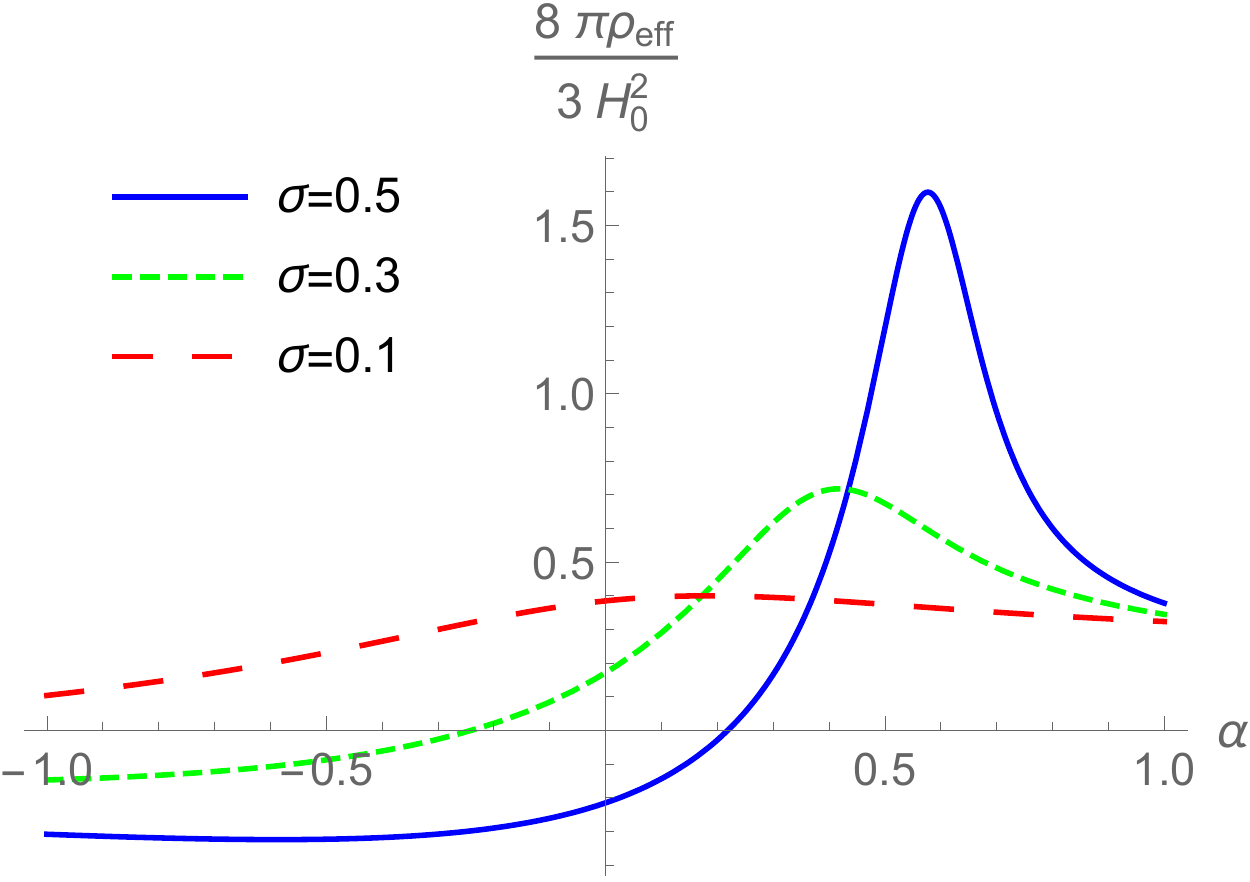}
\endminipage
\caption{Constant $\sigma$ profile plots of equation (\ref{wfc3}) for the hyperbolic tangent cosmological model $f=6H_0^2\left(\bar{T}+\sigma\bar{T}^\alpha\tanh\left(1/\bar{T}\right)\right)$. The solid blue curves show the $\sigma=0.5$ profiles, the dotted green curves show the $\sigma=0.3$ profiles and the dashed red curves show the $\sigma=0.1$ profiles. The left, middle and right panels correspond to $C=0$, $C=1$ and $C=10^4$ respectively.}
\label{fig: wtanh}
\end{figure*}

\begin{figure*}[!htb]
\minipage{0.32\textwidth}
  \includegraphics[width=\linewidth]{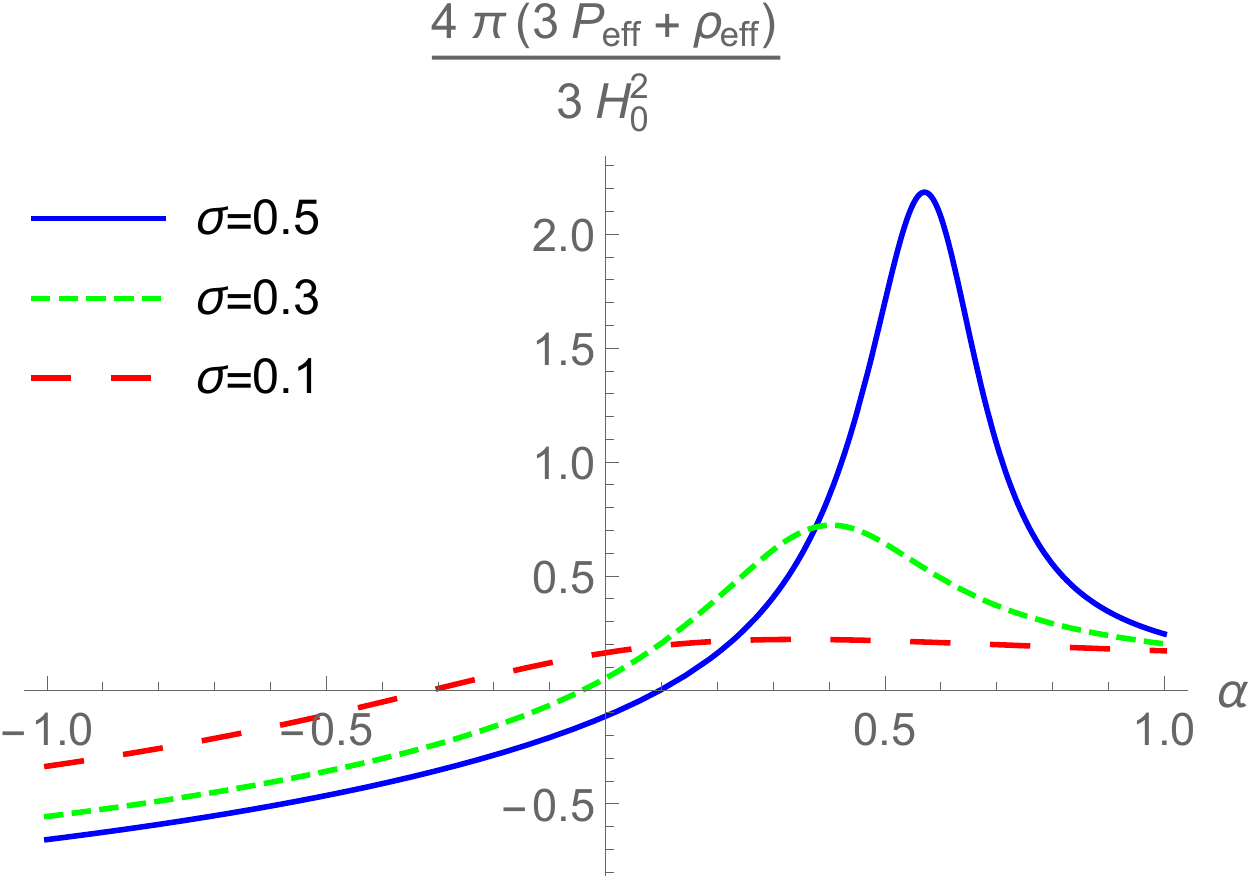}
\endminipage\hfill
\minipage{0.32\textwidth}
  \includegraphics[width=\linewidth]{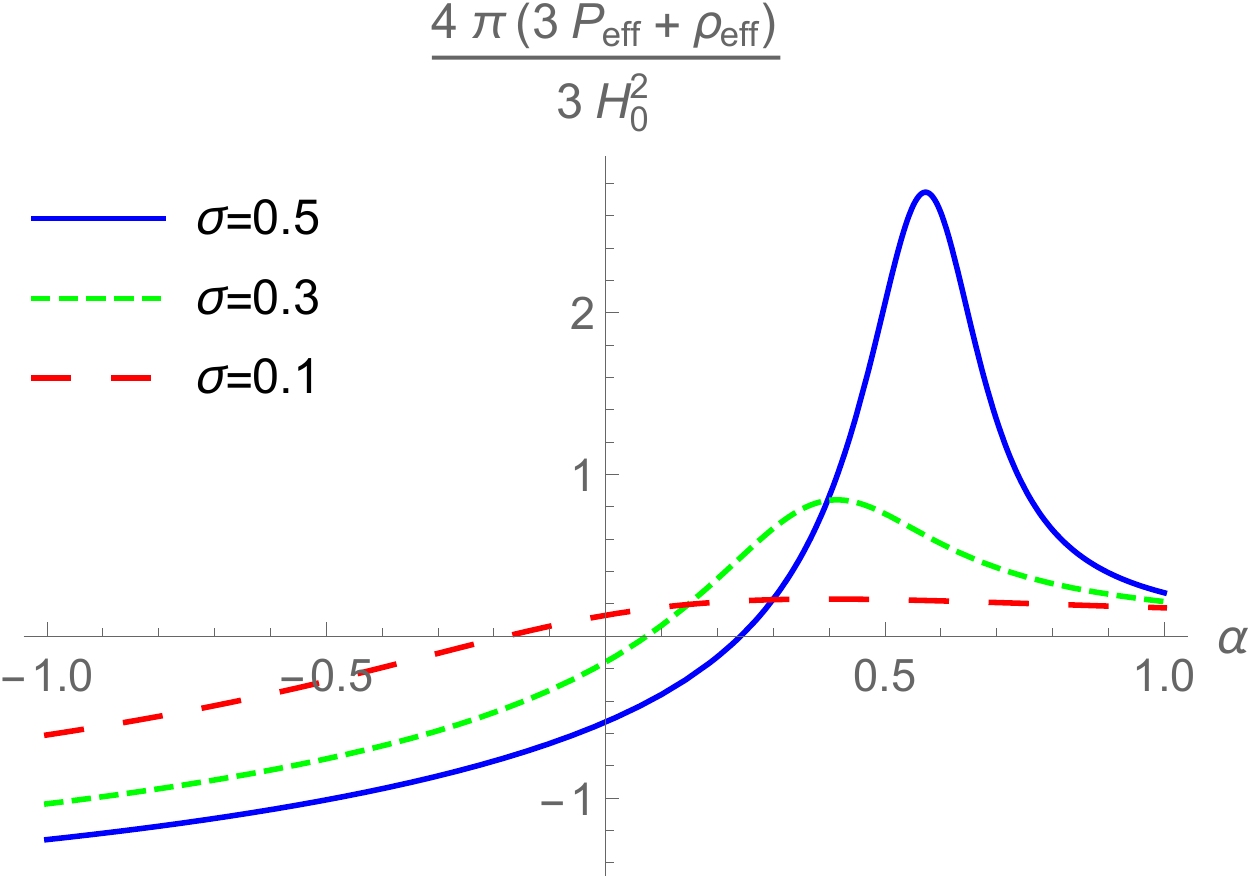}
\endminipage\hfill
\minipage{0.32\textwidth}%
  \includegraphics[width=\linewidth]{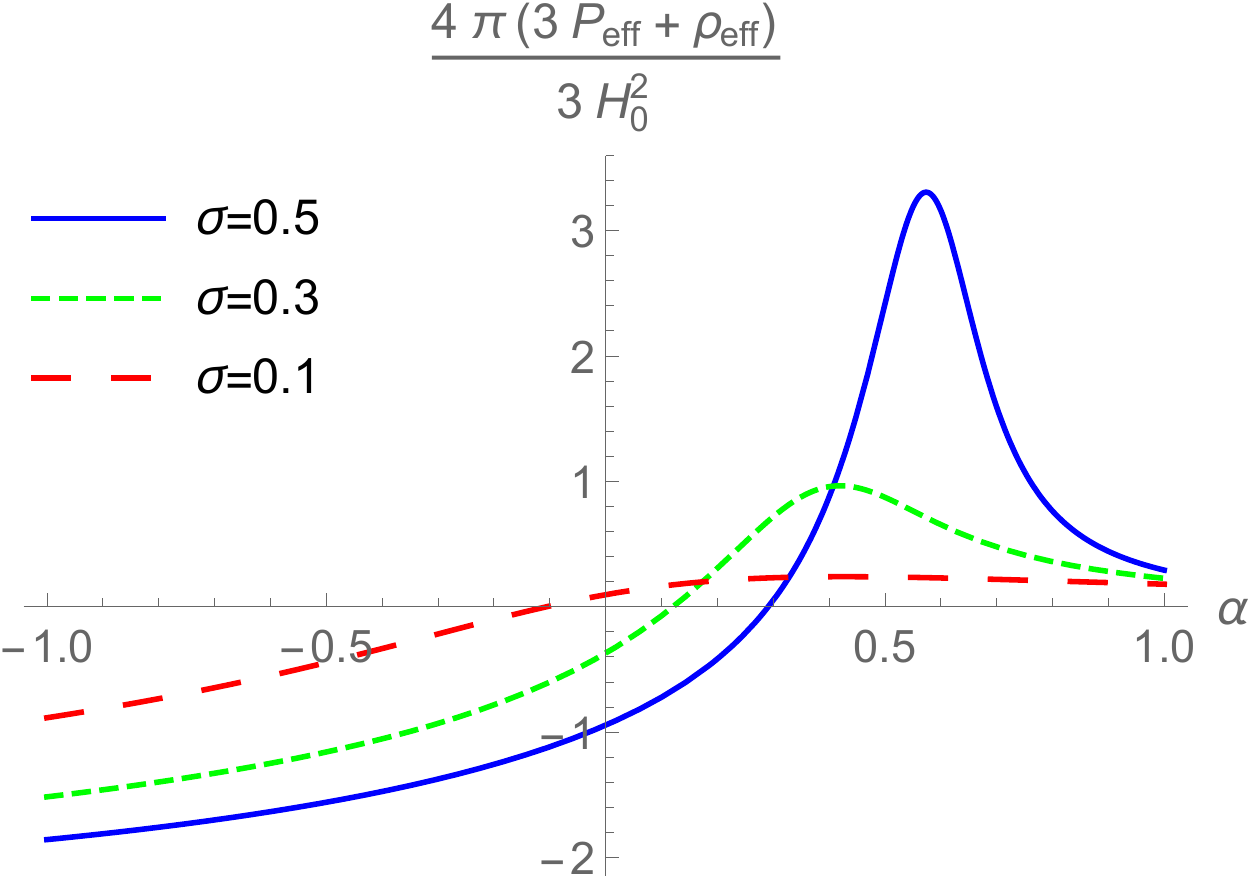}
\endminipage
\caption{Constant $\sigma$ profile plots of equation (\ref{sfc3}) for the hyperbolic tangent cosmological model $f=6H_0^2\left(\bar{T}+\sigma\bar{T}^\alpha\tanh\left(1/\bar{T}\right)\right)$. The solid blue curves show the $\sigma=0.5$ profiles, the dotted green curves show the $\sigma=0.3$ profiles and the dashed red curves show the $\sigma=0.1$ profiles. The left, middle and right panels correspond to $C=0$, $C=1$ and $C=10^4$ respectively.}
\label{fig: stanh}
\end{figure*}

The third and last cosmological model under consideration is the hyperbolic tangent model whose $f(T)$ Lagrangian is given above \cite{wu2010}. We have considered the $\alpha$ and $\sigma$ parameters lying in the $\left(\sigma,\alpha\right)\in\left[0,1/2\right]\times\left[-1,1\right]$ interval for which the $f(T)$ viability conditions described previously are satisfied. For this model, equation (\ref{feq2}) for the case of vanishing pressure reads
\begin{align}
&\frac{\textup{d}h}{\textup{d}z}=\frac{3}{2(1+z)}\Big[h^2+\sigma\left(2\alpha-1\right)h^{2\alpha}\tanh\left(1/h^2\right)-2\sigma h^{2\alpha-2}\sech^2\left(1/h^2\right)\Big]\Big[\Big(h+\sigma\alpha\left(2\alpha-1\right)\nonumber\\ &\times h^{2\alpha-1}\tanh\left(1/h^2\right)-\sigma\left(4\alpha-3\right)h^{2\alpha-3}\sech^2\left(1/h^2\right)-4\sigma h^{2\alpha-5}\sech^2\left(1/h^2\right)\tanh\left(1/h^2\right)\Big)\Big]^{-1}\ .
\end{align}
As done for the previous two cosmological models, we solve the above differential equation and obtain values for $h(z)$ for the various combinations of $\alpha$ and $\sigma$ values. From equation (\ref{weakonepara}), we find that the weak focusing condition for this cosmological model in the case of dust reads
\begin{align}
&\frac{8\pi\rho_{eff}}{3H_0^2}=\Big[1+\sigma\alpha h^{2\alpha-2}\tanh\left(1/h^2\right)-\sigma h^{2\alpha-4}\sech^2\left(1/h^2\right)\Big]^{-1}\Big[\Omega_{m_0}\left(1+z\right)^3\nonumber\\ &+\sigma\left(1-\alpha\right)h^{2\alpha}\tanh\left(1/h^2\right)-\sigma h^{2\alpha-2}\sech^2\left(1/h^2\right)-\frac{4h'\beta^2}{3\left(1+z\right)}\Big(\sigma\alpha\left(\alpha-1\right)h^{2\alpha-1}\nonumber\\ &\times\tanh\left(1/h^2\right)-2\sigma\left(\alpha-1\right)h^{2\alpha-3}\sech^2\left(1/h^2\right)\nonumber-2\sigma h^{2\alpha-5}\sech^2\left(1/h^2\right)\tanh\left(1/h^2\right)\Big)+h^2\nonumber\\ &+\sigma h^{2\alpha}\tanh\left(1/h^2\right)\Big]+\frac{\beta^2h^2}{\left(1+z\right)^2}\geq0\ .
\label{wfc3}
\end{align}
In addition, we find that, from equation (\ref{strongonepara}), the strong focusing condition for the hyperbolic tangent cosmological model in the case of dust reads
\begin{align}
&\frac{4\pi\left(\rho_{eff}+3P_{eff}\right)}{3H_0^2}=\Big[1+\sigma\alpha h^{2\alpha-2}\tanh\left(1/h^2\right)-\sigma h^{2\alpha-4}\sech^2\left(1/h^2\right)\Big]^{-1}\bigg[\frac{1}{2}\Omega_{m_0}\left(1+z\right)^3\nonumber\\ &-\sigma\left(1-\alpha\right)h^{2\alpha}\tanh\left(1/h^2\right)+\sigma h^{2\alpha-2}\sech^2\left(1/h^2\right)-2\Big(\sigma\alpha\left(\alpha-1\right)h^{2\alpha-1}\nonumber\tanh\left(1/h^2\right)\nonumber\\ &-2\sigma\left(\alpha-1\right)h^{2\alpha-3}\sech^2\left(1/h^2\right)-2\sigma h^{2\alpha-5}\sech^2\left(1/h^2\right)\tanh\left(1/h^2\right)\Big)\Big)\left(1+z\right)h'\nonumber\\ &-\frac{4h'\beta^2}{3\left(1+z\right)}\Big(\sigma\alpha\left(\alpha-1\right)h^{2\alpha-1}\tanh\left(1/h^2\right)-2\sigma\left(\alpha-1\right)h^{2\alpha-3}\sech^2\left(1/h^2\right)\nonumber\\ &-2\sigma h^{2\alpha-5}\sech^2\left(1/h^2\right)\tanh\left(1/h^2\right)\Big)\Big)+h^2+\sigma h^{2\alpha}\tanh\left(1/h^2\right)\bigg]+\frac{\beta^2h^2}{\left(1+z\right)^2}\geq0\ .
\label{sfc3}
\end{align}
As done for the previous two cosmological models, the energy density parameter $\Omega_{m_0}$ depends upon the cosmological parameters $\alpha$ and $\sigma$ through equation (\ref{feq1}) at the present day value of $z=0$. In particular
\begin{align}
\Omega_{m_0}&=h^2_0+\sigma\left(2\alpha-1\right)h_0^{2\alpha}\tanh\left(1/h^2_0\right)-2\sigma h_0^{2\alpha-2}\sech^2\left(1/h_0^2\right)\ .
\end{align}
Figure \ref{fig: rtanh} contains the region plots showing the satisfaction or violation of the weak and strong focusing conditions for the hyperbolic tangent cosmological model. The left, middle and right panels correspond to the choices $C=0$, $C=1$ and $C=10^4$ respectively. As was noticed for the previous two cosmological models, the higher the value of $C$ the bigger the area of the $\{\alpha,\,\sigma\}$ region for which the strong focusing condition is violated.  For this cosmological model, unlike the two previous models, we found that as the value for $C$ increases
regions for which the weak focusing condition is violated arise (see right panel in Figure \ref{fig: rtanh}). 
Figure \ref{fig: wtanh} shows the constant $\sigma$ profiles of equation (\ref{wfc3}). In the right panel, which corresponds to $C=10^4$, there are $\alpha$ and $\sigma$ combinations resulting in negative values for the effective energy density, $\rho_{eff}$. This demonstrates that, although the weak focusing condition may be satisfied for the case of a fundamental congruence ($C=0$), there may be values for $C$ for which  the weak focusing condition is not satisfied. Finally, Figure \ref{fig: stanh} shows constant $\sigma$ profiles of equation (\ref{sfc3}). There we see that, 
as the value of $C$ increases (from left to right  panels) and for a fixed $\sigma$, equation (\ref{sfc3}) seems to take smaller values which could be eventually negative. Therefore  the strong focusing condition may be violated.

%%%%%%%%%%%%%%%%%%%%%%%%%%%%%%%%%%%%%%%%%%%%%%%%%%
\section{Conclusions}
\label{Sec:8}
In this note, we studied the weak and strong focusing conditions for a one-parameter dependent congruence of timelike auto-parallels of the Levi-Civita connection in the context of $f(T)$ extended theories of teleparallel gravity. 
By noting that in the context of $f(T)$ theories of gravity, test particles follow auto-parallels of the Levi-Civita connection we made use of the torsion-free Raychadhuri equation in order to determine the inequalities to be satisfied so such conditions hold. Through the consideration of a spatially-flat Robertson-Walker space-time, we derived the weak and strong focusing conditions for a one-parameter, $C$, dependent congruence of timelike auto-parallels of the Levi-Civita connection. It was also shown that, in the case of a fundamental congruence, i.e., when $C=0$, the obtained focusing conditions reduced to those previously found in \cite{liu} as expected.
Once the general expressions were obtained, we also examined the obtained weak and strong focusing conditions for three viable bi-parametric $f(T)$ cosmological models when the only standard fluid present is dust matter. These cosmological models were of polynomial, exponential and hyperbolic tangent forms. Thus,  results presented in this work are expected to cover a wide phenomenology of more convoluted $f(T)$ models. 
For each cosmological model under consideration, we determined the Hubble parameter evolution in terms of the redshift using the $f(T)$ second Friedman equation, whereas the first one was used to determine the matter density parameter today as a function of the $f(T)$ parameters. Once the Hubble parameter had been solved for, the weak and strong focusing conditions were studied for a redshift value of $z=0$.
For each cosmological model considered, region plots were produced for the cases of $C=0$, $C=1$ and $C=10^4$. It was found that, for all three cosmological models, the increase in the value for $C$ resulted in the parameter-space region where the strong focusing condition is violated to be bigger. It was also found that, for the polynomial and exponential cosmological models, the weak focusing condition remained satisfied when the value for $C$ was increased. However, in the case of the hyperbolic tangent model it was found that a sufficient increase in the value for $C$ eventually resulted in the appearance of 
parameter-space regions for which the weak focusing condition is violated. Consequently, for this model this result would imply that although the weak focusing condition for a fundamental congruence is satisfied, one cannot conclude that the weak focusing condition would be satisfied for all possible auto-parallel curves of the Levi-Civita connection, i.e., for all possible values of $C$.

As a final comment, we wish to make a note on the focusing conditions that arise in the so-called $f(R)$ and $f(Q)$ theories of gravity. Analyses of the focusing conditions in the context of $f(R)$ theories can be found in \cite{Santos:2007bs,Capozziello:2018wul}. The authors of \cite{Santos:2007bs}
reported $f(R)$ models for which regions exist where the weak focusing condition is violated in addition to regions that allow for the weak focusing condition to be satisfied. The authors of \cite{Capozziello:2018wul} reported $f(R)$ models for which the weak as well as the strong focusing conditions can be satisfied or violated. The authors of \cite{Mandal:2020lyq} studied the focusing conditions in the context of $f(Q)$ theories of gravity. In \cite{Mandal:2020lyq}, the focusing conditions are studied using the present day values for the Hubble and deceleration parameters. There, $f(Q)$ models were reported where the weak as well as the strong focusing conditions could be satisfied or violated. Furthermore we note that, with regards to the $f(T,B)$ theories, one would obtain the $f(R)$ focusing conditions when $f(T,B)=f(-T,-B)$ while one would obtain the $f(T)$ focusing conditions reported here in the case where $f(T,B)=f(T)+B$. General focusing conditions in the $f(T,B)$ theories have been studied in \cite{Bhattacharjee:2020chn}.

In \cite{liu}, $f(T)$ models where the weak focusing condition could be satisfied or violated were reported. There, the authors made use of the present day values for the Hubble and deceleration parameters. In the present work, the $f(T)$ field equations for each considered model were solved and the solution $h(z)$ was substituted into the focusing conditions. In addition to this, we considered a one-parameter depend congruence, i.e., a general value for $C$. For each of the models considered, regions existed where the weak and strong focusing conditions could be satisfied. For each of these models, regions for which the strong focusing condition could be violated were found. Lastly, we note that for a value of $C=10^4$, regions exist for the hyperbolic tangent model where the weak focusing condition can be violated. These obtained results are specific to timelike auto-parallel curves of the Levi-Civita connection. For discussions on neighbouring null curves in $f(T)$ theories, the interested reader is directed to \cite{liu,Chakrabarti:2020ngy}.

Since all the mentioned modified gravities can describe the same kind of violations of the energy conditions one could use any of them to describe the accelerated expansion of the Universe without requiring Dark Energy. The choice of which one is more appropriate will need to be based on other aspects, such as the absence of ghost modes and instabilities, gravitational wave constraints, compatibility with fifth force and large scale-structure \cite{Capozziello:2018wul}.
%%%%%%%%%

\appendix

\section{Appendix: Derivation of the torsion-free Raychaudhuri equation}
\label{Appendix}
Let us derive here the Raychaudhuri equation for a smooth congruence of timelike curves generated by the Levi-Civita connection, $D$. First, let us introduce the physical quantities describing the expansion, the shear and the twist for such a congruence. Since the Levi-Civita connection, $D$, is nothing more than the affine connection $\nabla$ in the absence of torsion, we can define the \textit{torsion-free expansion}, $\overset{\circ}\theta$, the \textit{torsion-free shear}, $\overset{\circ}\sigma_{\alpha\beta}$, and the \textit{torsion-free twist}, $\overset{\circ}\omega_{\alpha\beta}$, for this congruence by replacing $\nabla$ in equations (\ref{re7})-(\ref{re9}) by $D$. Accordingly, the torsion-free expansion scalar, the torsion-free shear tensor and the torsion-free twist tensor can be written as 
\begin{align}
\overset{\circ}\theta &= h_\alpha^{\ \beta}D_\beta\xi^\alpha \ , \label{circtheta} \\ 
\overset{\circ}\sigma_{\alpha\beta} &= h^\sigma_{\ \alpha}h^\rho_{\ \beta}D_{(\sigma}\xi_{\rho)}-\frac{1}{3}\overset{\circ}\theta h_{\alpha\beta}\ , \label{circsigma} \\
%\end{align}
%and
%\begin{align}
\overset{\circ}\omega_{\alpha\beta} &=  h^\sigma_{\ \alpha}h^\rho_{\ \beta}D_{[\sigma}\xi_{\rho]} \ , \label{circtwist}
\end{align}
respectively \cite{wald4,kar,Capozziello:2001mq}. We note that we can relate the geometric, non-null torsion and torsion-free kinematic quantities through the following expressions
\begin{align}
\overset{\circ}\theta &= \theta + T_\beta\xi^\beta=\tilde\theta \ , \label{thetatotheta} \\
\overset{\circ}\sigma_{\alpha\beta} &= {\sigma}_{\alpha\beta} - \frac{1}{3}T_\rho\xi^\rho h_{\alpha\beta}  + h^\sigma_{\ \alpha}h^\rho_{\ \beta}K^\mu_{\ (\sigma\rho)}\xi_{\mu}=\tilde\sigma_{\alpha\beta} \  , \label{sigmatosigma} \\
\overset{\circ}\omega_{\alpha\beta} &= {\omega}_{\alpha\beta} + h^\sigma_{\ \alpha}h^\rho_{\ \beta}K^\mu_{\ [\sigma\rho]}\xi_\mu=\tilde\omega_{\alpha\beta}-\xi^\nu h^\sigma_{\ \alpha}h^\rho_{\ \beta}K_{\sigma\nu\rho} \ . \label{twisttotwist}
\end{align}
In order to obtain the torsion-free Raychaudhuri equation for a smooth congruence of timelike curves generated by the Levi-Civita connection, $D$, one can simply consider 
 equation (\ref{re42}) and assume there a vanishing torsion tensor, i.e., $T^\rho_{\ \alpha\beta}=0$. In such a case, the non-null torsion quantities $\theta$, $\sigma_{\alpha\beta}$, $\omega_{\alpha\beta}$ and $\tilde{R}_{\alpha\beta\sigma}^{\ \ \ \ \rho}$ reduce to the torsion-free counterparts $\overset{\circ}\theta$, $\overset{\circ}\sigma_{\alpha\beta}$, $\overset{\circ}\omega_{\alpha\beta}$ and ${R}_{\alpha\beta\sigma}^{\ \ \ \ \rho}$ respectively. Accordingly equation (\ref{re42}) becomes 
\begin{eqnarray}\label{re42lc}
\frac{\textup{d}\overset{\circ}\theta}{\textup{d}\lambda} = D_\alpha\left(\xi^\beta D_\beta\xi^\alpha\right) - \overset{\circ}\sigma_{\alpha\beta}\overset{\circ}\sigma\phantom{}^{\alpha\beta} + \overset{\circ}\omega_{\alpha\beta}\overset{\circ}{\omega}\phantom{}^{\alpha\beta}- \frac{1}{3}\overset{\circ}{\theta}\phantom{}^2 - {R}_{\alpha\beta}\xi^\alpha\xi^\beta \ . 
\end{eqnarray}
In Riemann-Cartan space-time, $\mathcal{U}_4$, the above expression can be used to describe a smooth congruence of timelike auto-parallels of the Levi-Civita connection, i.e., curves satisfying $\xi^\beta D_\beta\xi^\alpha=0$. By considering the tangent vector fields to be hypersurface orthogonal, we have $\overset{\circ}{\omega}_{ab}=0$ and thus, for a smooth congruence of timelike auto-parallels of the Levi-Civita connection, equation (\ref{re42lc}) becomes
\begin{align}\label{re42lcc}
\frac{\textup{d}\overset{\circ}\theta}{\textup{d}\tau} = -\overset{\circ}\sigma_{\alpha\beta}\overset{\circ}\sigma\phantom{}^{\alpha\beta} - \frac{1}{3}\overset{\circ}\theta\phantom{}^2 - {R}_{\alpha\beta}\xi^\alpha\xi^\beta \ ,
\end{align}
where $\tau$ is the proper time \cite{wald4}. 

% Equation (\ref{re42lcc}) is the Raychaudhuri equation used in the study of torsion-free theories of gravity when considering a smooth congruence of timelike geodesics \cite{wald4}. \\ \\

\acknowledgments
We thank Paulo Luz and Felipe Mena for insightful comments. The financial assistance of the National Research Foundation (NRF) towards this research is hereby acknowledged. Opinions expressed and conclusions arrived at, are those of the authors and are not necessarily to be attributed to the NRF. Authors acknowledge financial support from NRF Grants No.120390, Reference: BSFP 190-416431035, and No.120396, Reference: CSRP190405427545, and No 101775, Reference: SFH 1507-27131568. UKBV also acknowledges the UCT Postgraduate Funding Office for financial support through the Myer Levinson Scholarship and the Vice Chancellor Research Scholarship.
AdlCD also acknowledges financial support from Project No. FPA2014-53375-C2-1-P from the Spanish Ministry of Economy and Science, MICINN Project
No. PID2019-108655GB-I00, Project No. FIS2016-78859-P from the European Regional
Development Fund and Spanish Research Agency (AEI), and support from Projects Nos.
CA15117 and CA16104 from COST Action EU Framework Programme Horizon 2020. 
FJMT acknowledges financial support from the Erasmus+ KA107 Alliance4Universities programme and from the Van Swinderen Institute at the University
of Groningen. 

\bibliographystyle{JHEP}
\bibliography{references}

\end{document}